\documentclass[aps, reprint, pra, twocolumn, superscriptaddress,amsmath,amssymb, longbibliography,noeprint]{revtex4-2}
\usepackage{times}
\usepackage{amsmath}
\usepackage{amssymb}
\usepackage{xfrac}
\usepackage{dsfont}
\usepackage{graphicx}
\usepackage{float}
\usepackage{braket}
\usepackage{bbold}
\usepackage[normalem]{ulem}
\usepackage{bm}
\usepackage{bbm}
\usepackage{color}
\usepackage{xcolor}
\usepackage[colorlinks=true, urlcolor=blue, linkcolor=blue, citecolor=blue]{hyperref}
\usepackage{siunitx}

\def\ahfs{A_{\mathrm{hfs}}}
\def\ahfsf{\frac{A_{\mathrm{hfs}}}{2}}

\def\ii{{\rm i}} 
\def\rtwo{r^2_{m,F,F'}}
\def\stwo{s^2_{m,F,F'}}

\def\rscross{r_{m,F,F'}s_{m+2,F,F'}}
\def\lx{{\{\!\{}} 
\def\rx{{\}\!\}}} 
\def\rb{\mathbf{r}}

\def\cpar{\hat{c}_\parallel}
\def\cperp{\hat{c}_\perp}

\newcommand{\berkeleyphy}{Department of Physics, University of California, Berkeley, California 94720}
\newcommand{\CIQC}{Challenge Institute for Quantum Computation, University of California, Berkeley, California 94720}
\newcommand{\LBL}{Materials Sciences Division, Lawrence Berkeley National Laboratory, Berkeley, California 94720}
\newcommand{\columbia}{Department of Physics, Columbia University, New York, NY 10027}

\begin{document}

\title{State-insensitive wavelengths for light shifts and photon scattering from Zeeman states}

\author{Stuart J. Masson}
\affiliation{\columbia}

\author{Zhenjie Yan}
\affiliation{\berkeleyphy}
\affiliation{\CIQC}

\author{Jacquelyn Ho}
\affiliation{\berkeleyphy}
\affiliation{\CIQC}

\author{Yue-Hui Lu}
\affiliation{\berkeleyphy}
\affiliation{\CIQC}

\author{Dan M. Stamper-Kurn}
\affiliation{\berkeleyphy}
\affiliation{\CIQC}
\affiliation{\LBL}

\author{Ana Asenjo-Garcia}
\email[]{ana.asenjo@columbia.edu}
\affiliation{\columbia}

\begin{abstract}
Atoms are not two-level systems, and their rich internal structure often leads to complex phenomena in the presence of light. Here, we analyze off-resonant light scattering including the full hyperfine and magnetic structure. We find a set of frequency detunings where the induced atomic dipole is the same irrespective of the Zeeman state, and where two-photon transitions that alter the atomic state turn off. For alkali atoms and alkaline-earth ions, if the hyperfine splitting is dominated by the magnetic dipole moment contribution, these detunings approximately coincide. Therefore, at a given ``magical'' detuning, all Zeeman states in a hyperfine manifold behave almost identically, and can be traced out to good approximation. This feature prevents state decoherence due to light scattering, which impacts quantum optics experiments and quantum information applications.
\end{abstract}

\maketitle

\section{Introduction}

The simplest toy model of quantum matter is a two-level system. This theoretical concept finds its realization in different experimental platforms, ranging from trapped ions~\cite{Blatt12,Bruzewicz19} and neutral atoms~\cite{Hammerer10,Browaeys20}, to artificial two-level systems based on superconducting circuits~\cite{Kjaergaard20} or solid-state devices~\cite{Ren19,Garcia21}. In particular, trapped ions and neutral atoms are generally addressable at optical frequencies and are inherently identical, which makes them ideal building blocks of larger systems such as quantum computers~\cite{Ladd10,Preskill18}, quantum simulators~\cite{Georgescu14,Altman21}, and metrological devices~\cite{Cronin09,Ma11,Pezze18}. 

However, the internal structure of atoms -- with fine, hyperfine, and Zeeman levels -- prevents them from behaving as true two-level systems, {instead displaying} open transitions and complex dynamics involving many levels~\cite{Birnbaum06,Asenjo19,PineiroOrioli22}. This rich level structure allows for state preparation through optical pumping~\cite{Happer72}, the realization of scalable quantum information platforms~\cite{Daley08,Gorshkov09,Allcock21,NChen22,Wu22,Kang23,Lis23} and the exploration of spin models with larger-than-spin-1/2 particles~\cite{Sadler06,Klempt10,Hamley12,Luo17,Masson17,Zhiqiang17,Davis19,Periwal21}, {among many applications. Nevertheless, open transitions are in many cases undesired}, as they lead to qubit decoherence and heating. However, restricting atoms to a specific subset of levels is challenging, as it requires initial state preparation and either strict adherence to selection rules to prevent leakage out of a two-level subspace, or continual repumping (e.g., optically) to recover atoms that have left the desired level subspace.

Under certain conditions, however, atoms behave identically irrespective of their internal state. One example is that of state-independent, or ``magic'', trapping. Here, the ground and excited states of a transition are both trapped by light of a particular wavelength and polarization such that the induced ac Stark shifts are the same for both levels~\cite{Ye08}. {Magic wavelengths create potentials for both states that have a consistent frequency separation in the presence of amplitude noise, therefore minimizing heating and limiting decoherence. Atoms held in magic wavelength potentials enable the realization of higher precision optical atomic clocks and more efficient quantum information protocols~\cite{Katori03,Takamoto05,Cooper18,Norcia18PRX,Saskin19}.  Magic wavelengths were first found for clock transitions in strontium~\cite{Katori99,Katori03,Ludlow06} and since extended to other atomic clock candidates such as ytterbium~\cite{Porsev04,Barber08} and mercury~\cite{Hachisu08}. Magic wavelengths have also been found for optical transitions in cesium~\cite{McKeever03} and in alkaline-earth ions~\cite{Haffner03,Kaur15,Liu15PRL}. Moreover, magic potentials generated by multiple frequencies have been implemented for optical transitions in rubidium~\cite{Lacroute12,Hilton19,Will21}.}

{In this paper, we extend the concept of magic wavelengths to Zeeman states in the same ground-state hyperfine manifold. We find conditions for state-independent photon scattering, showing that at a particular ``magic detuning'' all Zeeman states scatter incoming light in exactly the same manner.} Such state-independent scattering requires two criteria to be met: the amplitude and polarization of the induced dipole must be identical for all {Zeeman} states, and the atom must not change its {Zeeman} state as it scatters. We show that these requirements yield three separate conditions for the polarizability tensor. For a large class of atomic species, these conditions are all met at approximately the same frequency. {This coalescence requires: (1) scattering via three excited states, (2) for the hyperfine splitting to be dominated by the nuclear magnetic dipole contribution, and (3) for the atomic spectrum not to feature other relevant lines.} We find optimal detunings to achieve state-independent optical responses of alkali atoms and alkaline-earth ions, and find that scattering for the most optimal species, $^{133}$Cs, can be state-independent at up to a part in $10^7$. We show this condition also implies that the ac Stark shift becomes state-independent, and discuss its relevance to cavity quantum electrodynamics (QED) experiments, {where this effect was recently demonstrated and utilized~\cite{Zhenjie23}.}

Our paper proceeds as follows. In Sec.~\ref{sec:tensor}, we introduce our formalism to describe the response of an atom to an external field, the generalized polarizability tensor. In Sec.~\ref{sec:conditions}, we introduce the conditions needed such that the atomic Zeeman state becomes irrelevant. In Sec.~\ref{sec:detunings}, we find the detunings that satisfy those conditions. In Sec.~\ref{sec:hyperfine} we show that, for atoms where the hyperfine structure is dominated by the nuclear magnetic dipole moment, these detunings coalesce to a single ``magic detuning''. In Sec.~\ref{sec:howtooptimize} we discuss how to optimize the detuning to minimize unwanted scattering, and then perform that optimization for alkali atoms and alkaline-earth ions in Sec.~\ref{sec:optimized}. We connect these results to the ac Stark shift in Sec.~\ref{sec:acstark}. In Sec.~\ref{sec:cavity}, we discuss its relevance to cavity QED experiments and, in particular, to the experiment performed in Ref.~\cite{Zhenjie23}.

\begin{figure*}
\includegraphics[width=\textwidth]{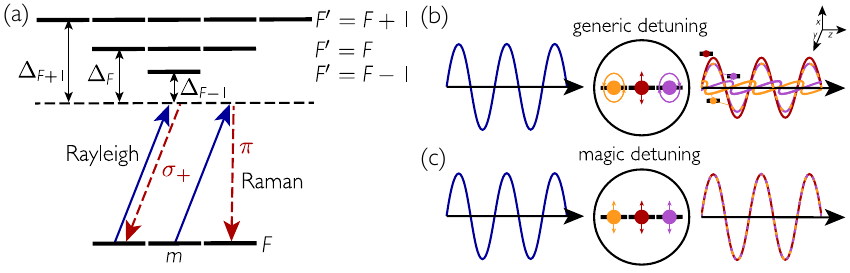}
\caption{Scattering from a {hyperfine-split line}. (a) {The atom is initially in a (ground state) hyperfine manifold $F$ (shown here as $F=1$ for simplicity, but generically $F\geq 1$).} The atom is then driven (blue) off-resonantly via the three lines with $F'\in\{F-1,F,F+1\}$, scattering light (red). Two types of scattering processes that can occur are depicted. The atom can Rayleigh scatter, returning to its original state, or Raman scatter, changing state during the process. (b,c) An atom in an $F=1$ state is driven by a linearly $x$-polarized field and scatters light. (b) For a generic detuning, an atom prepared in a superposition of states becomes entangled with the scattered field, as the polarization and amplitude of the field depend on the Zeeman state. (c) At a magic detuning, the generated field is independent of the {Zeeman} state.\label{fig:setup}}
\end{figure*}

\section{Generalized polarizability tensor\label{sec:tensor}}

We consider an atom driven by the electric field of monochromatic light. In the following, we describe the mathematical structure used in the rest of the paper. We discuss the level structure of the atom, how to characterize its response to the field, the coordinate basis used to perform the characterization, and how to evaluate the necessary dipole matrix elements.

\subsection{Atomic structure}

The ground state angular momentum of an atom is defined by its nuclear spin $I$ and its (ground state) total electron angular momentum $J$. Assuming $I + J \geq 1$, the ground state is split into different hyperfine manifolds labeled by $|I-J| \leq F\leq I+J$. In the absence of a magnetic field, each of these ground hyperfine manifolds has a $2F+1$ degeneracy due to its Zeeman sublevels. We consider light scattering via an excited state with total electron angular momentum $J'$. This state is also subject to hyperfine splitting, such that scattering is via three excited hyperfine manifolds $F'\in\,\{F-1,\,F,\,F+1\}$. This structure can be found in transitions such that $J'=J+1\, \,\forall\, F$, or such that $J'=J$ and $F \leq I+J-1$ or $J'=J-1$  and $F \leq I+J-2$. The necessary structure is most readily found in alkali atoms and singly ionized alkaline-earth ions  scattering from the $D_2$ line with $J'=J+1$. The simplest configuration of an $F=1$ hyperfine manifold is shown in Fig.~\ref{fig:setup}(a). The environment is assumed to be frequency selective -- e.g. a resonant optical cavity -- such that scattering from the $F'$ levels to other ground hyperfine manifolds can be neglected.

\subsection{Atomic response to an electric field}

We consider that the atom is driven by a monochromatic field of frequency $\omega$. The input field operator reads
\begin{equation}
\hat{\mathbf{E}}_{\mathrm{in}}(\mathbf{r}) =  \hat{\mathbf{E}}_\mathrm{in}^{(+)}(\mathbf{r})\,\mathrm{e}^{-i\omega t} + \mathrm{c.c},\label{eq:inputfield}
\end{equation}
where $\hat{\mathbf{E}}_\mathrm{in}^{(+)}$ is the  positive-frequency component of the field with polarization $\hat{\epsilon}$, and $\rb$ is the position of the atom. Here, for generality, the field is considered as an operator, but the treatment described in the rest of the manuscript is valid and unchanged if applied to a classical amplitude.

\color{black}

Under weak driving, such that the induced excited state population is negligible, the response of an atom in state $\ket{F,m_F}$ is captured by the polarizability tensor, i.e.,
\begin{align}
 &\alpha_{\mu\nu}(F,m_F,\omega) = \notag\\&-\sum\limits_{F',m_{F'}} \frac{\bra{F,m_F}d_\nu\ket{F',m_{F'}}\bra{F',m_{F'}}d_\mu\ket{F,m_F}}{\hbar\Delta_{F'}},\label{eq:polarizability}
\end{align}
where $\mu,\nu$ are polarization indices, $d_{\nu(\mu)}$ is the atomic dipole operator projected onto polarization $\nu (\mu)$, and $\Delta_{F'} = \omega-\omega_{F'}$ is the detuning of the drive field from the $F-F'$ resonance transition frequency. The above expression is derived under the rotating wave approximation, valid for $\Delta_{F'} \ll \omega_{F'}$.

The polarizability tensor encodes the dipole induced by the field on the atom and, therefore, also the field scattered by the atom. Specifically, an atom in state $\ket{F,m_F}$ at position $\rb$ subject to the input field defined in Eq.~\eqref{eq:inputfield} produces an output field at position $\rb'$ with $\chi$-polarized positive frequency component
\begin{equation}
\hat{\text{E}}^{(+)}_{\mathrm{out},\chi}(\rb') = \mu_0\omega^2\sum\limits_{\mu,\nu} G_{\chi\mu}(\rb',\rb,\omega) \alpha_{\mu\nu}(F,m_F,\omega)\hat{\text{E}}^{(+)}_{\mathrm{in},\nu}(\rb).
\end{equation}\color{black}
In the above equation, {$\mu_0$ is the permeability of free space}, $G_{\chi\mu}(\rb',\rb,\omega)$ is the Green's tensor of the electromagnetic environment {that describes the propagation of the output field from the atomic position to some position $\rb'$~\footnote{As the Green's tensor describes propagation of the electromagnetic field, its precise spectral and spatial form depends on the boundary conditions. For instance, the spatial dependence of the Green's tensor for the guided mode (of wavevector $k$) of a waveguide reads $G(z',z,\omega) \propto \exp({\ii k |z'-z|})$, while for a Fabry-P\'{e}rot cavity it reads $G(z',z,\omega) \propto \cos(k z)\cos(k z')$~\cite{Asenjo17PRA}, where in both cases we have assumed light propagation only along the $z$-axis.}, and $\hat{E}^{(+)}_{\mathrm{in}, \nu}(\rb)$ is the $\nu$-polarized positive frequency component of the input field}.

The standard form of the polarizability tensor does not describe all possible scattering processes, as it does not allow for the atom to change {from its initial Zeeman} state. The absence of state-changing transitions is usually justified by assuming they are strongly off resonant due to a large Zeeman splitting~\cite{Rosenbusch09}. However, in the absence of a large magnetic field, one should instead consider a more general map that includes different initial and final {Zeeman state}. Inspired by Eq.~\eqref{eq:polarizability}, we define the ``generalized polarizability tensor'' as
\begin{align}\label{poltens}
&\alpha^{nm}_{\mu\nu}(F,\omega) = \notag\\&- \sum\limits_{F',m_{F'}} \frac{\bra{F,n}d_\nu\ket{F',m_{F'}}\bra{F',m_{F'}}d_\mu\ket{F,m}}{\hbar\Delta_{F'}},
\end{align}
  where we have dropped the subscript $F$ from the ground {Zeeman} states, simply using $m$ and $n$ for the sake of clarity. 

In general, the state of the atom becomes correlated with the field, and the positive-frequency component of the field operator is related to the atomic coherence operators between {Zeeman states} as
\begin{align}
\hat{\text{E}}^{(+)}_{\mathrm{out},\chi}(\rb') = \mu_0\omega^2\sum\limits_{\nu,\mu,m,n}G_{\chi\mu}&(\rb',\rb,\omega)\,\alpha^{nm}_{\mu\nu}(F,\omega)\notag\\&\times \hat{\text{E}}^{(+)}_{ \mathrm{in},\nu}(\rb) \ket{F,m}\bra{F,n}.
\end{align}\color{black}
Elements of the generalized polarizability tensor that are diagonal in the atomic state (i.e., $n=m$), correspond to Rayleigh scattering processes and are simply given by the polarizability tensor of Eq.~\eqref{eq:polarizability}. Off-diagonal elements (i.e., $n\neq m$) correspond to Raman processes. Both these processes can generate correlations between the atom and scattered field, and so (even in the absence of Raman processes), an atom initially in a superposition of $m$ states will decohere as correlations develop. For the field to be independent of $\ket{m}\bra{n}$, we require the generalized polarizability tensor to be of the form
\begin{equation}
    \sum\limits_{m,n} \alpha^{nm}_{\mu\nu} (F,\omega)\ket{m}\bra{n} = \alpha_{\mu\nu} (F, \omega)\,\mathbb{1},
\end{equation}
i.e., for $\alpha^{nm}_{\mu\nu}$ to be diagonal in the {Zeeman} state (this is, only non-zero when $n=m$), and to be independent of $m$, {i.e., $\alpha^{mm}_{\mu\nu}(F,\omega) = \alpha_{\mu\nu}(F,\omega)~\forall~m$} such that the above sum over the atomic states becomes the identity. {Note that this condition means that for all {Zeeman} states we recover the conventional polarizability tensor of Eq.~\eqref{eq:polarizability}, with the additional constraint that it is independent of $m$. Thus, if a state-insensitive wavelength can be found, the polarizability tensor can be calculated by choosing any $m$ value in Eq.~\eqref{eq:polarizability}.} In the following, we use these criteria to find conditions on the dipole matrix elements, and find detunings at which the atomic state does not become correlated with the field.

\begin{figure*}[t!]
\includegraphics[width=\textwidth]{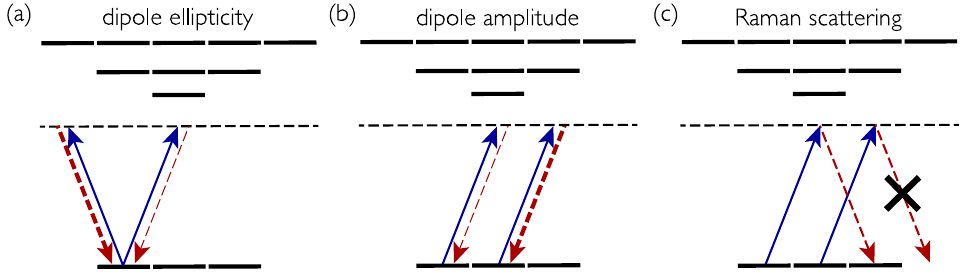}
\caption{Scattering processes that depend on the {Zeeman} state. (a) The ellipticity of the field may not be preserved in the induced dipole. Due to symmetry, states $m$ and $-m$ will respond in opposite ways, and thus this process is $m$ dependent. (b) The amplitude of the induced dipole may be different for different states. (c)~Scattering processes that change the {Zeeman} state can occur. Since not all states can undergo these processes, they are $m$ dependent.\label{fig:imperfections}}
\end{figure*}

\subsection{Choice of polarization basis}

The coordinate system can be arbitrarily chosen without loss of generality, and so we choose both the quantization axis of the atoms and the propagation direction of the drive to be $z$. Therefore, the polarization vector of the drive field lies in the $x$-$y$ plane~\footnote{Note that this choice of polarization is not the same as that in Ref.~\cite{Zhenjie23}, where it was assumed the (linear-polarized) drive was $\pi$-polarized. Here, we choose a different polarization axis because it is more convenient for the algebra to work in the $x-y$ plane for arbitrary elliptical polarization. Of course, the physics is unaffected by this choice, and our results will hold for any choice of polarization axis.}. Within this plane, it is convenient to work in a basis defined by the circular polarization unit vectors $\hat{\epsilon}_\pm = (\hat{\epsilon}_x \pm \ii \hat{\epsilon}_y) / \sqrt{2}$. Using this convention, we define a unit polarization vector parallel to that of the incoming light, and one perpendicular, of the form
\begin{equation}
    \hat{\epsilon}_\parallel = \hat{\epsilon}_+\cos\theta  +\hat{\epsilon}_- \sin\theta ,~~~~
  \hat{\epsilon}_\perp = \hat{\epsilon}_+\sin\theta \\ - \hat{\epsilon}_-\cos\theta.
\end{equation}
The angle $\theta$ encodes the ``ellipticity'' of the drive: for $\theta=\{0, \pi/2\}$, the field is purely circular, for $\theta=\pi/4$, the field is purely linear. This choice of basis requires setting the $x-y$ axes such that \begin{equation}
\hat{\epsilon}_\parallel = \hat{\epsilon}_x(\cos\theta + \sin\theta)/\sqrt{2} + \ii \hat{\epsilon}_y (\cos\theta -\sin\theta)/\sqrt{2},
\end{equation}
which is to say that the $x$-axis is defined to be identical to the real part of the input light polarization vector.

\subsection{Evaluating dipole matrix elements}

Due to selection rules, only three sets of dipole matrix elements can be non-zero: those corresponding to $\sigma_\pm$- and $\pi$-polarized transitions. We define the dipole matrix elements as [see Appendix~\ref{app:dme} for full expressions]
\begin{subequations}
\begin{align}
r_{m,F,F'} &= \bra{F,m} d_+ \ket{F',m+1}, \\
s_{m,F,F'} &= \bra{F,m} d_- \ket{F',m-1}, \\
t_{m,F,F'} &= \bra{F,m} d_\pi \ket{F',m}.
\end{align}
\end{subequations}
These selections rules can be used to eliminate the sums over $m_{F'}$ in the generalized polarizability tensor of Eq.~\ref{poltens}.

\section{Conditions for state independent scattering\label{sec:conditions}}

Here, we find conditions where the scattering response of the atom is independent of the atomic state. For the generalized polarizability tensor not to correlate the output light with the atomic state, it needs to be independent of the input and output atomic state. This requires Rayleigh scattering from all states to occur at the same amplitude, and for all states not to be able to Raman scatter.

\subsection{Rayleigh scattering}

The first {required} condition is that the atom Rayleigh-scatters (i.e., $m=n$) at a rate that is independent of {its Zeeman} state. For scattering into the parallel mode, we need $\alpha^{mm}_{\parallel\parallel}(F,\omega) = \alpha_{\parallel\parallel}(F,\omega)~\forall~m$. These terms are given by
\begin{align}
    \alpha&^{mm}_{\parallel\parallel}(F,\omega)= \notag\\& - \sum\limits_{F',m_{F'}} \frac{\bra{F,m}d_\parallel\ket{F',m_{F'}}\bra{F',m_{F'}}d_\parallel\ket{F,m}}{\hbar\Delta_{F'}}.
\end{align}
Since $\hat{\epsilon}_\parallel = \cos\theta \,\hat{\epsilon}_+ + \sin\theta \,\hat{\epsilon}_-$, the above equation can be written in terms of the dipole matrix elements as
\begin{align}
    \alpha^{mm}_{\parallel\parallel} &= - \sum\limits_{F'} \frac{r^2_{m,F,F'} + s^2_{m,F,F'}}{\hbar\Delta_{F'}} \notag \\&\;\;\;\;- \sum\limits_{F'}\frac{(\cos^2\theta - \sin^2\theta) \left(r^2_{m,F,F'} - s^2_{m,F,F'}\right)}{\hbar\Delta_{F'}}.
\end{align}
For linear polarization, $\cos\theta = \sin\theta$, and for the above expression to be state independent there is only one condition: that the sum of the two different-handed terms is independent of $m$. This condition {restricts the amplitude of the induced dipole to be identical} for all states. 

For arbitrary elliptical polarization, the difference between the two terms must also be independent of $m$. The dipole matrix elements have the symmetry
\begin{equation}
    r^2_{m,F,F'} = s^2_{-m,F,F'}.
\end{equation}
Therefore, the difference between the two different-handed terms is zero for the $m=0$ state. For this difference to be independent of $m$, it must be zero for all $m$. This condition enforces that the induced dipole preserves the ellipticity of the input field.

We also require that Rayleigh scattering into the perpendicular mode occurs at the same rate for all $m$. These elements can be written as
\begin{equation}
    \alpha^{mm}_{\perp\parallel}(F,\omega) = - \sum\limits_{F'} \frac{\bra{F,m}d_\perp\ket{F',m_{F'}}\bra{F',m_{F'}}d_\parallel\ket{F,m}}{\hbar\Delta_{F'}}.
\end{equation}
Note that $\alpha^{mm}_{\pi\parallel} = 0 \,\,\forall \,m$ because the atomic state must change to conserve spin when scattering $\pi$-polarized light. As above, this expression can be written in terms of dipole matrix elements as
\begin{equation}
    \alpha^{mm}_{\perp\parallel}(F,\omega) = - \sin\theta\cos\theta \sum\limits_{F'} \frac{ \left(r_{m,F,F'}^2 - s_{m,F,F'}^2\right)}{\hbar\Delta_{F'}}.
\end{equation}
For pure circularly-polarized input light ($\theta=\{0,\pi/2\}$) this condition is trivially met, as the above equation describes the difference in ellipticity of the input field and the induced dipole, and light with a given circular polarization cannot induce a dipole of opposite handedness. For input light with arbitrary polarization, the condition is akin to that above: the difference between the two different-handed terms must be zero for all $m$.

\subsection{Raman scattering}

The {Zeeman} state should not change in the scattering process. Not all states can Raman scatter both polarizations of light, as shown in Fig.~\ref{fig:imperfections}(c), and so the only possible way to eliminate $m$ dependence from Raman scattering is for it to be zero. Therefore,
\begin{equation}
    \alpha^{nm}_{\mu\parallel} = - \sum\limits_{F'} \frac{\bra{F,n}d_\mu\ket{F',m_{F'}}\bra{F',m_{F'}}d_\parallel\ket{F,m}}{\hbar\Delta_{F'}}=0
\end{equation}
for all $n\neq m$ and all output polarizations. Due to selection rules, only two types of Raman processes can occur, with associated polarizabilities
\begin{subequations}
    \begin{align}
        \alpha^{m+2,m}_{-+} &= -\sum\limits_{F'} \frac{r_{m,F,F'}s_{m+2,F,F'}}{\hbar\Delta_{F'}}, \\
        \alpha^{m+1,m}_{\pi+} &= -\sum\limits_{F'} \frac{r_{m,F,F'}t_{m+1,F,F'}}{\hbar\Delta_{F'}}.
    \end{align}
\end{subequations}
Exploiting the generalized polarizability tensor symmetry
\begin{equation}
    \alpha^{-n,-m}_{\nu\mp} = \alpha^{n,m}_{\nu\pm},
\end{equation}
it is easy to see that processes that decrease the magnetic number are also zero when the two conditions above are met.

\subsection{Summary}

In summary, three conditions on the dipole matrix elements need to be satisfied for state-independent light-scattering. If they are all met at the same drive frequency for all $m$, then the generalized polarizability tensor is independent of the {Zeeman} state and the field will not be correlated with the atomic state. The conditions are summarized below:
\begin{enumerate}
    \item \textbf{Preserved dipole ellipticity.} The difference between the different-handed dipole matrix elements must be zero. This implies
    \begin{equation}
    \sum\limits_{F'} \frac{r^2_{m,F,F'} - s^2_{m,F,F'}}{\Delta_{F'}} = 0.\label{eq:cond1}
    \end{equation}
    {We denote the detuning at which this condition is met $\Delta_\perp$, as the induced dipole has no component perpendicular to the input field.} 
    \item \textbf{Equal dipole amplitude.} The sum of the different-handed dipole matrix elements must be independent of $m$. This implies
    \begin{equation}
    \sum\limits_{F'} \frac{r^2_{m,F,F'} + s^2_{m,F,F'}}{\Delta_{F'}} = f(F,F').\label{eq:cond2}
    \end{equation}
    {We denote the detuning at which the condition is met $\Delta_\parallel$,  as the parallel component of the induced dipole has the same amplitude for all states.} 
    \item \textbf{Zero Raman scattering.} Raman processes must not occur. This is guaranteed if: 
    \begin{subequations}
     \label{cond3}
    \begin{align}
        \sum\limits_{F'} \frac{r_{m,F,F'}s_{m+2,F,F'}}{\Delta_{F'}} = 0, \label{cond3a}\\
        \sum\limits_{F'} \frac{r_{m,F,F'}t_{m+1,F,F'}}{\Delta_{F'}}=0.\label{cond3b}
    \end{align}
    \end{subequations}
    We define the detunings that satisfy these equations as $\Delta_{c,\pi}$ respectively.
\end{enumerate}
In the following sections, we find detunings that separately satisfy each of these conditions.

\section{Detunings to meet state-independent scattering conditions\label{sec:detunings}}

\subsection{Preserved dipole ellipticity\label{sec:ellipticity}}

Here, we find the detuning that satisfies Eq.~\eqref{eq:cond1}, which ensures that the induced dipole has the same polarization as the input field. {In our considered setup, the sum has three terms with $F' \in \{F-1,F,F+1\}$}. The set of three detunings $\{\Delta_{F'}\}$ are not independent from each other, and two of them can be defined as differences from the other. We define the detuning from the central $F'=F$ line, i.e., $\Delta_F \equiv \Delta$, such that $\Delta_{F\pm1} = \Delta + \zeta_{\pm1}$, where $\zeta_{\pm1} = \omega_{F} - \omega_{F\pm1}$ is the hyperfine splitting. By direct evaluation of the dipole matrix elements, the above sum becomes (see Appendix~\ref{app:dperp} for details)
\begin{equation}
\sum_{l=-1}^1 (-1)^l\frac{(2 + l(2F+1)-l^2)(2F+1)-2l}{\Delta_\perp + |l|\,\zeta_{l}} \lx F+l\rx = 0,\label{eq:d2ellipticitycondition}
\end{equation}
where $\lx F+l\rx$ denotes the square of a Wigner 6-j symbol of the form
\begin{equation}
\lx F+l\rx \equiv \begin{Bmatrix} J & J' & 1 \\ F+l & F & I \end{Bmatrix}^2,    
\end{equation}
where $J$ and $J'$ are the electron angular momentum quantum numbers of the ground and excited states, respectively, and $I$ is the nuclear angular momentum quantum number. Importantly, Eq.~\eqref{eq:d2ellipticitycondition} contains no $m$ dependence, and so $\Delta_\perp$ is the same for all $m$ states. This occurs because $r^2_{m,F,F'} - s^2_{m,F,F'}$ is always linear in $m$, and so it simply factors out. The above equation has two trivial solutions, $\Delta_\perp \rightarrow \pm \infty$. These solutions simply describe that if the probe light is infinitely far detuned then all $m$ states have no response, trivially making them respond in the same way. There are also two non-trivial solutions that can be found by rearranging the expression into a quadratic equation in $\Delta_\perp$. {For the specific case of scattering on a $D_2$ line, they are real}. Details on the form of the solutions and their derivation are provided in Appendix~\ref{app:dperp}.

\subsection{Equal dipole amplitude\label{sec:amplitude}}

Here we find the detuning that satisfies Eq.~\eqref{eq:cond2} and ensures that the induced dipole parallel to the input field has the same amplitude for all $m$ states. Following the same procedure as above, we expand the sum in Eq.~\eqref{eq:cond2} as three terms with $F' \in \{F-1,F,F+1\}$ and then expand the dipole matrix elements. In this case we find that $r^2_{m,F,F'} + s^2_{m,F,F'}$ has two contributions: a scalar term and a term that is quadratic in $m$. For the scattering rate to be $m$ independent we thus require the sum over the terms quadratic in $m$ to be zero. As above, since the $m$ dependence is the same for every term, it simply factors out, and any solution that is found works for all $m$ states. The detuning at which the above condition is met for all $m$ states is thus the solution to
\begin{equation}
\sum_{l=-1}^1 (-1)^l\frac{(2-l^2)(2F+1)-l}{\Delta_{\parallel} + |l|\zeta_{l}} \lx F+l\rx = 0.\label{eq:amplitudecondition}
\end{equation}
This equation has two trivial solutions, $\Delta_\parallel \rightarrow \pm \infty$. To find potential non-trivial solutions, we again rearrange the terms into a quadratic equation. In {the case of scattering on a $D_2$ line}, the coefficient for the $\Delta_\parallel^2$ term is zero, and thus the condition has only one non-trivial solution. Details on the form of the solution and its derivation are provided in Appendix~\ref{app:dpar}.

\subsection{Zero Raman scattering\label{sec:raman}}

\subsubsection{Circular scattering}

Here, we find a detuning that satisfies Eq.~\eqref{cond3a}. Substituting in expressions for the dipole matrix elements (see Appendix~\ref{app:dme}) reduces the expression to Eq.~\eqref{eq:amplitudecondition}, and so Raman processes involving two circular photons are zero at $\Delta_\parallel$.

\subsubsection{$\pi$ scattering}

Here, we find a detuning that satisfies Eq.~\eqref{cond3b}. By substituting in expressions for the dipole matrix elements as before we arrive at the condition
\begin{widetext}
\begin{align}
\sum\limits_{l=-1}^1 (-1)^l \frac{(2F+1)(2m+2) + l (F(2F+2) - m - 1) - l^2 (2F+1)(m+1)}{\Delta_\pi + |l|\,\zeta_l} \lx F+l\rx = 0.
\end{align}
\end{widetext}
{This expression has two parts: one proportional to $m$ and one independent of $m$. The $m$-dependent terms take a familiar form, that of Eq.~\eqref{eq:amplitudecondition}}. The Raman rate thus becomes $m$ independent at $\Delta_\parallel$. The $m$ independent part is zero at a third detuning condition $\Delta_\pi$. {For scattering on a $D_2$ line}, the expression for $\Delta_\pi$ can be simplified using the zero combination of the three Wigner 6-j symbols we found in the derivation of $\Delta_\parallel$ [Eq.~\eqref{eq:importantzero} in App.~\ref{app:dpar}], and is given as
\begin{equation}
\Delta_\pi = \frac{\zeta_{+1}\lx F\rx}{\lx F+1\rx - \lx F\rx}.
\end{equation}

\section{Approximating the hyperfine splitting~\label{sec:hyperfine}}

The hyperfine splitting can be expressed in terms of a sum over different nuclear multipole moments~\cite{Schwartz55,Allegrini22}. Typically, the largest contribution is given by the magnetic dipole moment interaction, described by the magnetic dipole hyperfine constant $A_{\mathrm{hfs}}$, with the largest corrections arising from the electric quadrupole and magnetic octopole moments, quantified by $B_{\mathrm{hfs}}$ and $C_{\mathrm{hfs}}$ respectively. 

If the magnetic dipole moment is the sole contributor (or by far the most dominant one), the hyperfine splitting reads
\begin{equation}
\delta_{\mathrm{hfs}} = \ahfsf \left[F'(F'+1) - I(I+1) - J(J+1)\right].
\end{equation}
The hyperfine splitting between excited states is thus
\begin{subequations}
 \begin{gather}
 \zeta_{+1} = -A_{\mathrm{hfs}}(F+1), \\
 \zeta_{-1} = A_{\mathrm{hfs}}F.
 \end{gather}
\end{subequations}
Substituting these values into the expressions for $\Delta_\perp, \Delta_\parallel$ and $\Delta_\pi$ we find that all three of our detunings have a common root (full details of this derivation are provided in Appendix~\ref{app:hype}). This brings us to the key finding of our paper: for atoms scattering on a $D_2$ line, there is a detuning where the generalized polarizability tensor does not correlate the {Zeeman} state with the field at all.

\section{Optimization of the state-insensitive detuning\label{sec:howtooptimize}}

We have shown that, for cases where hyperfine structure results only from the nuclear magnetic dipole moment, the hyperfine structure interval rule guarantees that all three state-independence criteria can be met at a single detuning.  However, this interval rule does not strictly hold when higher-order nuclear moments are included.  In this more general situation, and as shown in Table I, we no longer find a single detuning at which all three criteria apply.  Nevertheless, to the extent that higher-order nuclear moments contribute only weakly to hyperfine structure, the three state-independent criteria are all  approximately fulfilled at the same detuning.

Here, somewhat \textit{ad hoc}, we derive a detuning at which the conditions for state-independent scattering are all nearly satisfied. We approach this problem by considering the distance between the generalized polarizability tensor from an ``ideal'' one, defined as $\tilde{\alpha}^{nm}_{\mu\parallel}(F,\omega)$ with entries
\begin{subequations}
\begin{align}
    \tilde{\alpha}^{mm}_{\parallel\parallel}(F,\omega) &= \alpha_{\parallel\parallel}^{00}(F,\omega)\;\forall\;m, \\
    \tilde{\alpha}^{nm}_{\mu\parallel}(F,\omega) &= 0\;\forall\;n\neq m, \mu\neq\parallel.
\end{align}
\end{subequations}
The ``magic distance'' is obtained from the square of the (normalized) Frobenius norm of the difference between these two tensors, i.e.,
\begin{equation}
    M(F,\Delta) = \sum\limits_{n,m,\mu} \left|\frac{\tilde{\alpha}^{nm}_{\mu\parallel}(F,\omega) - \alpha^{nm}_{\mu\parallel}(F,\omega)}{\alpha^{00}_{\parallel\parallel}(F,\omega)} \right|^2.\label{eq:magicness}
\end{equation}
Since we use the polarization of the input field to define our basis, the only relevant parts of the generalized polarizability tensor are those with polarization parallel to the input field. As such, we only concern ourselves with the output polarization index, and initial and final atomic state indices. For species where the hyperfine spin is half-integer, there is no $m=0$ state, and the normalization is instead performed with the $m=1/2$ state. We consider that minimizing $M(F, \Delta)$ provides a ``nearly state-independent scattering'' detuning. 

\begin{figure}[t!]
    \includegraphics[width=.5\textwidth]{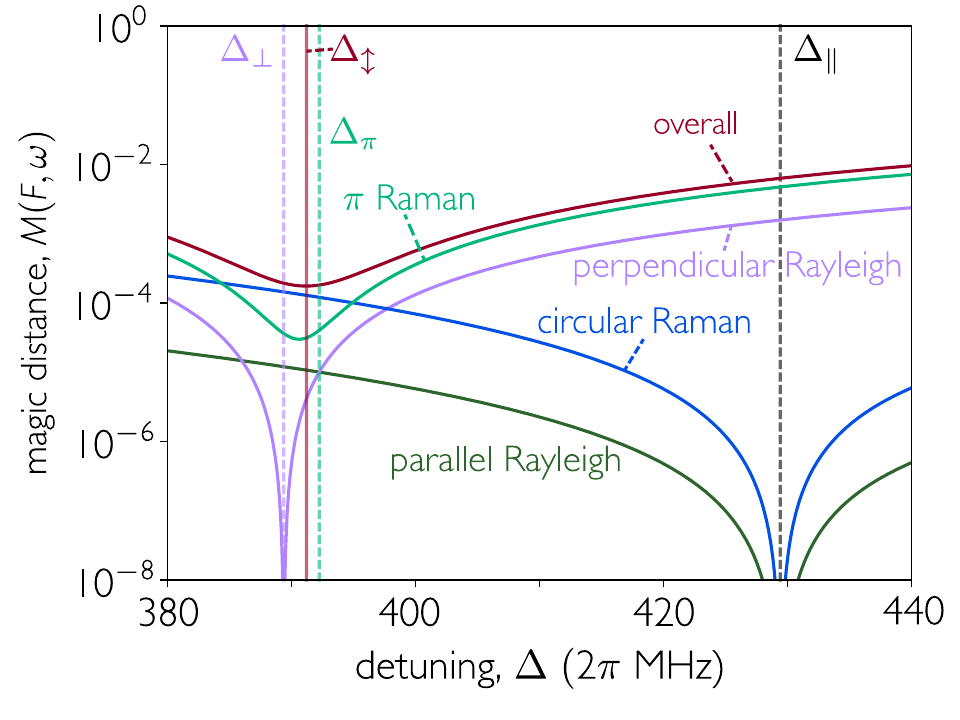}
    \caption{Optimization to find the magic detuning. The overall value of the magic distance defined by Eq.~\eqref{eq:magicness}, is shown along with the different contributions to that value. Vertical dashed lines show the detunings that meet each condition, while the solid vertical line shows the optimized detuning. The example shown here is a $^{87}$Rb atom in the $F=1$ line driven with linear light.\label{fig:optimizeprocess}}
\end{figure}

 \begin{table*}[t!]
 \begin{center}
 \begin{tabular}{| c | c | c | c | c | c | c | c | c | c | c | c | c |}
 \hline
 Atom & $F$ & $\zeta_{+1}$ ($2\pi$ MHz) & $\zeta_{-1}$ ($2\pi$ MHz) &  $B_{\mathrm{hfs}} / A_{\mathrm{hfs}}$ & $\Delta_\perp$ ($2\pi$ MHz) & $\Delta_\parallel$ ($2\pi$ MHz) & $\Delta_\pi$ ($2\pi$ MHz) & $\Delta_\updownarrow$ ($2\pi$ MHz) & $M(F,\Delta_\updownarrow)$\\\hline
 $^{6}$Li~\cite{Allegrini22} & 3/2 & 2.9 & -1.8 & 0.087 & 0.58, 2.38 & 2.41 & 2.32 & 2.39 & $6.7\times10^{-4}$ \\\hline
 $^{7}$Li~\cite{Shimizu87} & 1 & 6.0 & -2.9 & 0.052 & 1.45, -15.0 & -15.5 & -15.0 & -15.0 & $3.0\times10^{-5}$ \\\hline
 $^{7}$Li~\cite{Shimizu87} & 2 & 9.4 & -6.0 & 0.052 & 1.55, 9.11 & 8.92 & 9.40 & 9.05 & $9.4\times10^{-4}$ \\\hline
 $^{23}$Na~\cite{Yei93} & 1 & -34.3 & 15.8 & 0.147 & -8.0, 85.1 & 93.8 & 85.8 & 85.5 & $1.7\times10^{-4}$ \\\hline
 $^{23}$Na~\cite{Yei93} & 2 & -58.3 & 34.3 & 0.147 & -9.4, -53.4 & -50.3 & -58.3 & -52.4 & $7.9\times10^{-3}$ \\\hline
 $^{40}$K~\cite{Falke06} & 7/2 & 33.3 & -24.2 & 0.45 & 3.5, -73.0 & -77.8 & -74.0 & -73.4 & $5.7\times10^{-4}$ \\\hline
 $^{40}$K~\cite{Falke06} & 9/2 & 44.1 & -33.3 & 0.45 & 3.9, 60.6 & 57.6 & 64.1 & 60.1 & $3.0\times10^{-3}$ \\\hline
 $^{85}$Rb~\cite{Das08} & 2 & -63.4 & 29.4 & 1.03 & -7.7, 140.6 & 227.6 & 147.9 & 143.9 & $7.2\times10^{-3}$ \\\hline
 $^{85}$Rb~\cite{Das08} & 3 & -120.6 & 63.4 & 1.03 & -13.4, -118.9 & -97.3 & -150.8 & -113.9 & $6.0\times10^{-2}$ \\\hline
 $^{87}$Rb~\cite{Ye96} & 1 & -156.9 & 72.2 & 0.148 & -36.4, 389.4 & 429.4 & 392.2 & 391.2 & $1.8\times10^{-4}$\\\hline
 $^{87}$Rb~\cite{Ye96} & 2 & -266.7 & 156.9 & 0.148 & -42.8, -244.3 & -229.9 & -266.7 & -239.5 & $8.0\times10^{-3}$ \\\hline
 $^{133}$Cs~\cite{Gerginov03} & 3 & -201.29 & 151.22 & -0.00981 & -25.20, 453.04 & 452.36 & 452.90 & 452.99 & $3.0\times10^{-7}$ \\\hline
 $^{133}$Cs~\cite{Gerginov03} & 4 & -251.09 & 201.29 & -0.00981 & -25.12, -352.05 & -352.50 & -351.53 & -352.13 & $1.9\times10^{-6}$ \\\hline
 \end{tabular}
 \end{center}
 \caption{Properties of alkali atoms, values of detunings that meet certain conditions, and optimized detunings for minimal magic distance. Only one line is presented for $^6$Li as, due to its nuclear spin $I=1$, scattering from the $F=1/2$ level is only via two $F'$ levels. Two isotopes of potassium, $^{39}$K and $^{41}$K, are also omitted, as the hyperfine structure of the $D_2$ line is close to degenerate and not known to sufficient accuracy to perform these calculations~\cite{Falke06,Das08}.\label{tab:alkaliconditions}}
 \end{table*}

The dependence of the magic distance with frequency for $^{87}$Rb is shown in Fig.~\ref{fig:optimizeprocess}. Imperfections in the generalized polarizability tensor due to Rayleigh scattering into the perpendicular and parallel mode are minimized at $\Delta_\perp$ and $\Delta_\parallel$, respectively. Raman scattering of circular light is also minimized at $\Delta_\parallel$, while Raman scattering of $\pi$-polarized light is minimized close to $\Delta_\pi$. Note that this does not occur exactly at $\Delta_\pi$ because that detuning is only where the state-independent fraction of the Raman scattering is zero, as the $m$ dependent part is minimized at $\Delta_\parallel$, and so minimizing $\pi$-polarized Raman scattering is a compromise between these two values. This is also the reason why the primary contribution of imperfections to the generalized polarizability tensor is $\pi$-polarized Raman scattering. As such, the overall optimal detuning for this species lies relatively close to $\Delta_\pi$. Nevertheless, the magic distance is relatively flat across a fairly broad range of detunings, and small perturbations from the optimal detuning do not drastically increase the magic distance.

\section{Optimized detuning\label{sec:optimized}}

In this Section, we calculate the relevant detunings for different species of atoms and atomic ions. We focus on alkali atoms and singly-ionized alkaline-earth atoms. In both cases, the single outer electron provides a simple and clean spectrum, and the values of $J$ and $J'$ are always such that the above analysis is valid. We optimize for linear polarization, yielding an optimized detuning we define as $\Delta_\updownarrow$. It should be noted that the optimized detuning does depend on the polarization of the input field, but the impact is small. The optimized detuning for a circular input field is generally within 1~MHz of $\Delta_\updownarrow$, and for elliptical polarizations the optimal detuning varies smoothly between these values.

\subsection{Alkali atoms}

We first consider alkali atoms. These encompass species with both integer and half-integer hyperfine spin, as well as a variety of nuclear spin. Table~\ref{tab:alkaliconditions} shows the detuning conditions $\Delta_{\perp,\parallel,\pi}$ for experimentally relevant alkalis, as well as the optimized detuning $\Delta_\updownarrow$ and the magic distance at that detuning.

\begin{figure}[b!]
\includegraphics[width=0.5\textwidth]{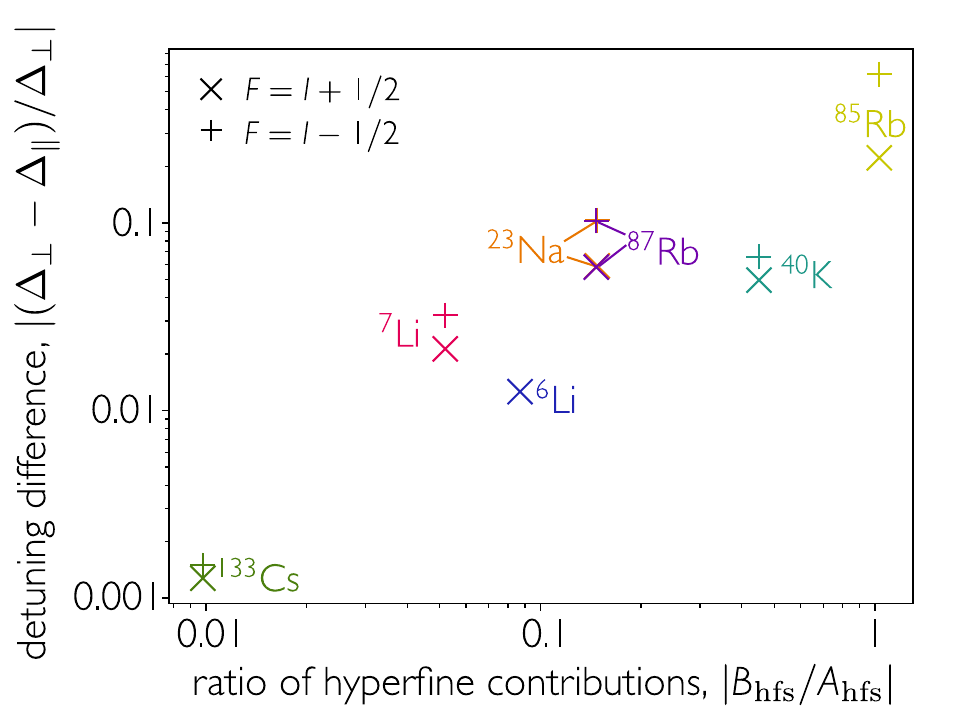}
\caption{The normalized difference between $\Delta_\perp$ and $\Delta_\parallel$, as a function of the ratio of (electric quadrupole vs magnetic dipole) hyperfine contributions, $|B_\mathrm{hfs}/A_{\mathrm{hfs}}|$. $\Delta_\perp$ is chosen as the solution closest to $\Delta_\parallel$. Generically, the two detunings are close to one another if the relative contribution of the electric quadrupole moment is very small. Data is shown for ground state hyperfine manifolds $F=I+1/2$ (+ symbols) and $F=I-1/2$ ($\times$ symbols).\label{fig:magicness}}.
\end{figure}

The detuning conditions share a similar root for all considered species. The most dissimilar are found for $^{85}$Rb, where the $F=2$ line has roots almost 80~MHz apart. The closest are found for $^{133}$Cs, where they are within 1~MHz of each other. As anticipated, this is directly related to the contributions to the hyperfine splitting, as $^{85}$Rb and $^{133}$Cs have the highest and lowest relative electric quadrupole contributions, respectively. Figure~\ref{fig:magicness} shows that, generically, the relative detuning difference decreases together with the relative electric quadrupole contribution. The two species that do not perfectly fit this trend are the two fermionic isotopes $^6$Li and $^{40}$K, where the difference is smaller than the general trend. For these species, the different nuclear spin means the electric quadrupole shift contribution is of a different magnitude even for the same $B_{\mathrm{hfs}}$. Since $^{23}$Na and $^{87}$Rb have the same ratio of hyperfine contributions and the same nuclear spin, they also have the same detuning difference. For all species, the detuning difference is smaller for the hyperfine level $F=I+1/2$ than for $F=I-1/2$.

The usefulness of these magic wavelengths is likely greatly reduced for lithium experiments, as the detuning is found very close to a resonance. For $^6$Li, the detuning conditions are met less than 1~MHz from resonance with the $F'=1/2$ excited state. For $^7$Li, they are met around 6~MHz and 10~MHz from excited state resonances for the $F=1$ and $F=2$, respectively. Therefore, to meet the criterion of essentially no atomic excitation, one would need to use extremely weak laser power. Furthermore, there will be much higher sensitivity to frequency differences between {Zeeman states}.

\begin{table*}
 \begin{center}
 \begin{tabular}{| c | c | c | c | c | c | c | c | c | c | c | c | c |}
 \hline
 Ion & $F$ & $\zeta_{+1}$ ($2\pi$ MHz) & $\zeta_{-1}$ ($2\pi$ MHz) &  $B_{\mathrm{hfs}} / A_{\mathrm{hfs}}$ & $\Delta_\perp$ ($2\pi$ MHz) & $\Delta_\parallel$ ($2\pi$ MHz) & $\Delta_\pi$ ($2\pi$ MHz) & $\Delta_\updownarrow$ ($2\pi$ MHz) & $M(F,\Delta_\updownarrow)$\\\hline
 $^{43}$Ca$^+$~\cite{Nortershauser98} & 3 & 122.0 & -88.1 & 0.223 & 14.8, -272.6 & -282.3 & -274.5 & -273.3 & $1.5\times10^{-4}$ \\\hline
 $^{43}$Ca$^+$~\cite{Nortershauser98} & 4 & 159.9 & -122.0 & 0.223 & 15.8, 216.4 & 210.3 & 223.9 & 215.4 & $1.0\times10^{-3}$ \\\hline
 $^{85}$Sr$^+$~\cite{Buchinger90} & 4 & 275.0 & -315.9 & -4.08 & 36.3, -657.8 & -503.2 & -605.0 & -642.2 & $0.023$ \\\hline
 $^{85}$Sr$^+$~\cite{Buchinger90} & 5 & 154.1 & -275.0 & -4.08 & 15.0, 423.5 & 846.3 & 213.2 & 450.2 & $0.199$ \\\hline
 $^{87}$Sr$^+$~\cite{Buchinger90} & 4 & 198.4 & -203.0 & -2.46 & 24.0, -461.0 & -381.1 & -436.5 & -453.4 & $0.010$ \\\hline
 $^{87}$Sr$^+$~\cite{Buchinger90} & 5 & 157.0 & -198.4 & -2.46 & 14.4, 323.8 & 439.7 & 235.5 & 336.1 & 0.069 \\\hline
 $^{89}$Sr$^+$~\cite{Buchinger90} & 2 & 171.9 & -77.9 & 1.07 & 20.5, -380.2 & -643.7 & -401.1 & -389.7 & $8.0\times10^{-3}$ \\\hline
 $^{89}$Sr$^+$~\cite{Buchinger90} & 3 & 331.9 & -171.9 & 1.07 & 36.6, 324.4 & 263.2 & 414.9 & 310.2 & 0.065 \\\hline
 $^{91}$Sr$^+$~\cite{Buchinger90} & 2 & 163.3 & -115.0 & -0.234 & 28.5, -384.5 & -365.4 & -381.0 & -383.0 & $2.1\times10^{-4}$ \\\hline
 $^{91}$Sr$^+$~\cite{Buchinger90} & 3 & 200.6 & -163.3 & -0.234 & 27.5, 273.2 & 271.2 & 275.8 & 272.8 & $7.4\times10^{-5}$ \\\hline
 $^{135}$Ba$^+$~\cite{Villemoes93} & 1 & -167.0 & 54.0 & 0.522 & -27.9, 403.7 & 920.2 & 417.5 & 413.0 & $3.8\times10^{-3}$ \\\hline
 $^{135}$Ba$^+$~\cite{Villemoes93} & 2 & -398.0 & 167.0 & 0.522 & -56.3, -294.9 & -233.2 & -398.0 & -272.6 & 0.118 \\\hline
 $^{137}$Ba$^+$~\cite{Villemoes93} & 1 & -161.9 & 34.7 & 0.727 & -18.4, 382.6 & -1216.0 & 404.8 & 398.4 & 0.011 \\\hline
 $^{137}$Ba$^+$~\cite{Villemoes93} & 2 & -474.1 & 161.9 & 0.727 & -60.9, -314.9 & -221.3 & -474.1 & -283.9 & 0.251 \\\hline
 $^{221}$Ra$^+$~\cite{Neu88} & 2 & 681.1 & -1136.2 & -60.9 & 231.9, -1946.4 & -1073.6 & -1589.2 & -1833.0 & 0.084 \\\hline
 $^{221}$Ra$^+$~\cite{Neu88} & 3 & -1001.8 & -681.1 & -60.9 & -227.8, 623.9 & 734.9 & -1252.3 & N/A & N/A \\\hline
 $^{223}$Ra$^+$~\cite{Neu88} & 1 & 751.8 & -808.3 & 15.3 & 368.7, -2060.2 & -1224.5 & -1879.5 & -1965.4 & 0.025 \\\hline
 $^{223}$Ra$^+$~\cite{Neu88} & 2 & -1034.3 & -751.8 & 15.3 & -270.0, 720.6 & 795.2 & -1034.3 & N/A & N/A \\\hline
 \end{tabular}
 \end{center}
 \caption{Properties of singly-ionized alkaline earth ions and values of detunings that meet certain conditions, and optimized detunings for minimal magic distance. All integer hyperfine spin isotopes of alkaline-earth atoms have no nuclear spin, and therefore no hyperfine splitting, while $^{133}$Ba$^+$ cannot have a magic detuning due to its nuclear spin of $I=1/2$.  For the higher $F$ level for both radium isotopes, the optimization procedure for linear polarization finds no minima between $\Delta_\perp$ and $\Delta_\pi$, such that it instead finds the trivial $\Delta \rightarrow \infty$ solutions, and so no optimized detuning is obtained.\label{tab:alkalineionconditions}}
 \end{table*}

\subsection{Alkaline-earth ions}

For singly-ionized alkaline-earth ions, the hyperfine splitting contributions are generally not dominated by the magnetic dipole moment. For many ions, the optimized detuning is thus less magic than for the majority of alkali atoms, as shown in Table~\ref{tab:alkalineionconditions}. Two exceptions are $^{43}$Ca$^+$ and $^{91}$Sr$^+$, where the optimized detuning should provide close to magic behavior. For many ions, the electric quadrupole contribution is actually larger than the magnetic dipole contribution. This is best exemplified by the two radium isotopes, where for some choices of $F$ and input polarization, the optimization procedure instead finds the trivial $\Delta \rightarrow \infty$ solutions rather than any local minima.

\section{Applications\label{sec:applications}}

\subsection{Magic wavelength for {hyperfine manifolds}: state-independent ac Stark shift\label{sec:acstark}}

The ac Stark shift on a level $\ket{F,m_F}$ induced by the input light is readily given by
\begin{equation}
 \Delta E(F,m,\omega)= -\sum\limits_{\mu,\nu} \mathrm{Re}\left[\alpha^{mm}_{\mu\nu}(F,\omega)\right] E_\mu^{(-)}(\rb) E_\nu^{(+)}(\rb),
\end{equation}
{where $E_\mu^{(-)}(\rb) = \braket{\hat{E}_\mu^{(-)}(\rb)}$ is the expectation value of the field}. Unraveling the polarizability in terms of the dipole matrix elements, the shift reduces to
\begin{align}
 \Delta E(F,m,\omega) &= \sum\limits_{F'} \frac{|E_0^{(+)}|^2}{\Delta_{F'}} \left[ \left(r^2_{m,F,F'} + s^2_{m,F,F'}\right) + \right.\notag\\
 &\left.\left(\cos^2\theta - \sin^2\theta\right)\left(r^2_{m,F,F'} - s^2_{m,F,F'}\right)\right],
\end{align}
which implies that the ac Stark shift is also independent of $m$ for $\Delta_\parallel=\Delta_\perp$. As expected, the detuning that offers the closest to a magic wavelength depends on the polarization of the field.

The conditions we found above directly relate to the traditional description of the ac Stark shift in terms of scalar, vector, and tensor shifts. Our condition for preserved dipole ellipticity makes the vector Stark shift zero, while the condition for equal dipole amplitudes removes the $m^2$ dependence of the tensor shift.

While the values for the optimized detuning we find here are too close to resonance to efficiently trap atoms without significant atomic saturation and heating, the magic detuning can be useful in other contexts. For instance, the state independent ac Stark shift is translated to state-insensitive mechanical forces, potentially allowing for optomechanical applications where the {Zeeman} state can be traced out. The magic detuning could then be used to explore optomechanical driven-dissipative phase transitions~\cite{Baumann10,Baumann11,Klinder15}.

\subsection{State-independent scattering in cavity QED\label{sec:cavity}}

One consequence of the magic detuning is the ability to perform cavity and waveguide QED experiments without the need for careful state preparation. In waveguide QED, emitters are strongly coupled to a continuum of confined propagating modes~\cite{Vetsch10,Solano17}. Frequency selection to eliminate scattering to the other hyperfine ground state could be achieved with a photonic crystal waveguide~\cite{Goban15}. One complication is that the confinement of the field allows for a longitudinal polarization component, such that the coordinate system considered above is incomplete. However, Rayleigh scattering of $\pi$-polarized light must also be state-independent at the magic detuning due to symmetry arguments.

{Our predictions for state-independent light scattering were experimentally confirmed in Ref.~\cite{Zhenjie23}. In this experiment, a tweezer array of up to eight single rubidium atoms were placed in a cavity and driven with linearly polarized light. The atoms were able to scatter light into two degenerate cavity modes with polarization parallel or perpendicular to the initial drive light.} Generically, this is a complicated problem involving atoms scattering drive photons into  {both polarization modes of} the cavity, exchanging photons with the cavity, and changing their {Zeeman} state in the process. However, by operating at the magic detuning, one can eliminate the {Zeeman} state, reducing the Hamiltonian to an effective drive of the cavity and a dispersive shift to its frequency. In this experiment, scattering at the magic detuning enabled the observation of superradiant and subradiant scattering from an atomic array into the cavity. {Operation at the magic detuning guarantees that the induced atomic dipoles have the same polarization as the drive light and that only Rayleigh scattering occurs. The former was confirmed by measurements of the polarization of the cavity output light, while the latter was confirmed by the scaling of the scattered field with atom number.} A full derivation of the Hamiltonian in this limit is provided in Appendix~\ref{app:cavity}.

\section{Conclusions\label{sec:conclude}}

In conclusion, we have shown the generic existence of a ``magic detuning'' for scattering on the $D_2$ line of alkali atoms and alkaline-earth ions. At this detuning, a driven atom will respond in the same manner regardless of its {Zeeman state}, to very good approximation. Therefore, the complex level structure of a single hyperfine {manifold} driven off-resonantly via three excited manifolds reduces to that of an off-resonantly driven two-level atom. This eliminates the need of {Zeeman} state preparation in  quantum optics and atomic physics experiments. At the magic detuning, a prepared superposition of {Zeeman} states is not altered by the trapping light, nor does it decohere due to correlations with the scattered light. 

The magic detuning strictly occurs only in the absence of dc magnetic fields. Generally, Zeeman shifts must be small in comparison to the frequency range in which the response is close to magic. The magic condition also requires scattering to be via three excited hyperfine manifolds. Therefore, the nuclear spin has to be $I\geq1$. Scattering via {only two excited hyperfine manifolds} cannot be magic. The necessary conditions for scattering on {such a $D_1$ type line} can be met independently, but they do not coincide (see Appendix~\ref{app:d1line} for details). Moreover, we have neglected scattering to other hyperfine ground levels assuming that the environment is frequency selective (such as, for instance, a cavity). If the environment is not frequency selective, such as in free space, the lifetime of atoms would be limited by the rate of spontaneous emission to the other {ground state hyperfine manifold}, as spontaneous emission to both hyperfine {manifolds} cannot be state-independent at the same detuning. Repopulation of the desired hyperfine level is possible through a repump field, which can be trivially magic if the hyperfine splitting between the two hyperfine levels is much larger than that between the excited levels, such that the pump is effectively infinitely detuned. However, spontaneous emission and repumping will still decohere the atomic state. This limits the usefulness of the magic detuning in free space to situations where strictly maintaining coherence between the {Zeeman} states is not crucial.

The level to which the magic detuning conditions hold depends on the origin of the hyperfine splitting of the excited states. If the magnetic dipole dominates over the electric quadrupole (and higher order moments), the approximation is very good.  While we have focused on single-valence electron atoms and ions, the same physics may hold for the multi-electron case. To find {state-insensitive frequencies for Zeeman states}, one must find a combination of the quantum numbers that meets the necessary criterion given in Eq.~\eqref{eq:importantzero} in App.~\ref{app:dpar}. One has to be careful with multi-valence electron atoms, where the spectrum is more cluttered and hyperfine splitting is more complicated as it must include configuration interactions~\cite{Schwartz55,Fritzsche02}. However, as evidenced by the case of $^{85}$Rb, where the electric quadrupole contribution to the hyperfine splitting is of similar magnitude to that of the magnetic dipole, the magic detuning is very robust to other sources of hyperfine splitting. Therefore, if the hyperfine splitting is still given by the magnetic dipole contribution to even a broad approximation, the two-level picture may still hold. The presence of a magic detuning may also be interesting in the context of molecules~\cite{Zelevinsky08}, where magic conditions for trapping have been found for rotational states~\cite{Bause20,Guan21}. However, the presence of a cluttered spectrum due to rotational and vibrational states may prove problematic.
\\

\textbf{Acknowledgments-} We thank Luis Orozco and Francis Robicheaux for helpful comments. We acknowledge support from the AFOSR (Grant No.\ FA9550-1910328 and Young Investigator Prize Grant No.\ 21RT0751), from ARO through the MURI program (Grant No.\ W911NF-20-1-0136), from DARPA (Grant No.\ W911NF2010090), from the NSF (QLCI program through grant number OMA-2016245, and CAREER Award No.\ 2047380), and from the David and Lucile Packard Foundation. J.H. acknowledges support from the National Defense Science and Engineering Graduate (NDSEG) fellowship.

\onecolumngrid

\appendix

\section{Evaluating the dipole matrix elements~\label{app:dme}}

Using the Wigner-Eckart theorem, the dipole matrix elements for the atomic transitions are given by
\begin{equation}
    \bra{F,m}d_\nu\ket{F',m_{F'}} = \langle J ||\mathbf{d} || J' \rangle (-1)^{2F'+m_F+J+I}\sqrt{(2F+1)(2J+1)(2F'+1)}\begin{Bmatrix} J & J' & 1 \\ F' & F & I \end{Bmatrix} \begin{pmatrix} F & 1 & F' \\ m& q & -m- q \end{pmatrix},
\end{equation}
where $q \in \{-1,0,+1\}$ corresponds to photons with polarization $\nu \in \{\sigma_-,\pi,\sigma_+\}$, respectively. It is convenient to work in units such that $\langle J||\mathbf{d}||J'\rangle(-1)^{2F'+J+I}\sqrt{(2F+1)(2J+1)}=1$, as this is a constant for all transitions from a given hyperfine manifold. In these units, we define dipole matrix elements as
\begin{subequations}
\begin{align}
r_{m,F,F'} &= (-1)^{m}\sqrt{2F'+1}\begin{Bmatrix} J & J' & 1 \\ F' & F & I \end{Bmatrix} \begin{pmatrix} F & 1 & F' \\ m& 1 & -m- 1 \end{pmatrix}, \\
s_{m,F,F'} &= (-1)^{m}\sqrt{2F'+1}\begin{Bmatrix} J & J' & 1 \\ F' & F & I \end{Bmatrix} \begin{pmatrix} F & 1 & F' \\ m& -1 & -m+ 1 \end{pmatrix}, \\
t_{m,F,F'} &= (-1)^{m}\sqrt{2F'+1}\begin{Bmatrix} J & J' & 1 \\ F' & F & I \end{Bmatrix} \begin{pmatrix} F & 1 & F' \\ m& 0 & -m\end{pmatrix},
\end{align}
\end{subequations}
where $J$ is the total electron angular momentum quantum number, $I$ is the nuclear spin quantum number, and $\{~\}$ and $(~)$ represent Wigner 6-j and 3-j symbols respectively.

\subsection{Expressions for Wigner 3-j symbols\label{app:dme3j}}

The Wigner 3-j symbols can be calculated using the Racah formula. They are given by
\begin{subequations}
\begin{gather}
\begin{pmatrix} F & 1 & F' \\ m& 1 & -m- 1 \end{pmatrix} = (-1)^{2F+m-F'+1}\sqrt{\frac{2(F+F'-1)!}{(F'-F+1)!(F-F'+1)!(F+F'+2)!} \frac{(F- m)! (F'+ m+1)!}{(F+ m)!(F'- m-1)!}}, \\
\begin{pmatrix} F & 1 & F' \\ m& -1 & -m+ 1 \end{pmatrix} = (-1)^{F+m}\sqrt{\frac{2(F+F'-1)!}{(F'-F+1)!(F-F'+1)!(F+F'+2)!} \frac{(F+ m)! (F'- m+1)!}{(F- m)!(F'+ m-1)!}}.
\end{gather}
\end{subequations}

For $\pi-$polarized transitions, it is necessary to write expressions for each $F'$ separately. The Wigner 3-j symbols are
\begin{subequations}
\begin{gather}
\begin{pmatrix} F & 1 & F+1 \\ m& 0 & -m\end{pmatrix} = (-1)^{F-1+m} \sqrt{\frac{2(F+1-m)(F+1+m)}{(2F+1)(2F+2)(2F+3)}}, \\
\begin{pmatrix} F & 1 & F+1 \\ m& 0 & -m\end{pmatrix} = (-1)^{F-1+m} \frac{2m}{\sqrt{2F(2F+1)(2F+2)},}\\
\begin{pmatrix} F & 1 & F-1 \\ m& 0 & -m\end{pmatrix} = (-1)^{F+m} \sqrt{\frac{2(F+m)(F-m)}{(2F-1)2F(2F+1)}}.
\end{gather}
\end{subequations}

\subsection{Expressions for Wigner 6-j symbols}

Similarly, the Wigner 6-j symbols can also be calculated by the Racah formula. For the $D_1$ and $D_2$ lines respectively, they are given as
\begin{subequations}
\begin{align}
\begin{Bmatrix} \frac{1}{2} & \frac{1}{2} & 1 \\ F' & F & I \end{Bmatrix} &= \left(-1\right)^{F'+F+1} \sqrt{\frac{\left(F'-F+1\right)!\left(F-F'+1\right)!\left(F'+F+2\right)\left(F'+F+1\right)\left(F'+F\right)}{6\left(F+I+\frac{3}{2}\right)\left(F+I+\frac{1}{2}\right)\left(F'+I+\frac{3}{2}\right)\left(F'+I+\frac{1}{2}\right)}} \\
\begin{Bmatrix} \frac{1}{2} & \frac{3}{2} & 1 \\ F' & F & I \end{Bmatrix} &= \left(-1\right)^{F'+I+3/2}\left(I+F'+\frac{5}{2}\right)! \notag\\&\times\sqrt{\frac{\left(F'+\frac{3}{2}-I\right)!\left(I-F+\frac{3}{2}\right)!\left(F'+F-1\right)!}{12\left(I+F+\frac{3}{2}\right)\left(I+F+\frac{1}{2}\right)\left(F'+I-\frac{3}{2}\right)!\left(F'-F+1\right)!\left(F-F'+1\right)!\left(F+I+\frac{5}{2}\right)!\left(F'+F+2\right)!}}
\end{align} 
\end{subequations}

\section{Derivation of $\Delta_\perp$\label{app:dperp}}

In Sec.~\ref{sec:ellipticity}, we find a solution to
\begin{equation}
\sum\limits_{F'} \frac{\rtwo - \stwo}{\Delta_{F'}} = 0.\label{appeq:perpcond}
\end{equation}
Using the analytic expressions for the dipole matrix elements (see Appendix~\ref{app:dme}), this condition becomes
\begin{equation}
\sum\limits_{F'} \frac{(2F'+1) \begin{Bmatrix} J & J' & 1 \\ F' & F & I \end{Bmatrix}^2}{\Delta_{F'}} \left[\begin{pmatrix} F & 1 & F' \\ m & 1 & -m - 1 \end{pmatrix}^2 - \begin{pmatrix} F & 1 & F' \\ m & -1 & -m + 1 \end{pmatrix}^2 \right] = 0.
\end{equation}
We can then evaluate the Wigner 3-j symbols (see Appendix~\ref{app:dme3j} for expressions) to find
\begin{equation}
\sum\limits_{F'} 
\frac{1}{\Delta_{F'}}\frac{2(2F'+1) (F+F'-1)!\begin{Bmatrix} J & J' & 1 \\ F' & F & I \end{Bmatrix}^2}{(F'-F+1)!(F-F'+1)!(F+F'+2)!} \left[\frac{(F - m)! (F' + m+1)!}{(F + m)!(F' - m-1)!} - \frac{(F + m)! (F'- m+1)!}{(F- m)!(F'+ m-1)!} \right] = 0.
\end{equation}
For all possible values of $F' \in \{F-1,F,F+1\}$ the expression in the square brackets is always linear in $m$, such that all $m$ dependence can be eliminated. The solution to Eq.~\eqref{appeq:perpcond} is thus $m$ independent.

As in the main text, we denote the detuning at which induced ellipticity goes to zero as $\Delta_\perp$. We focus on the $D_2$ line, so we define this detuning from the central excited line $F'=F$, such that we make the substitutions $\Delta_{F}=\Delta_\perp$ and $\Delta_{F\pm1} = \Delta_\perp + \zeta_{\pm1}$. Our condition then becomes
\begin{equation}
\frac{2F(2F+3)}{\Delta_\perp + \zeta_{+1}} \begin{Bmatrix} J & J' & 1 \\ F+1 & F & I \end{Bmatrix}^2 
- \frac{4F+2}{\Delta_\perp}\begin{Bmatrix} J & J' & 1 \\ F & F & I \end{Bmatrix}^2
+ \frac{(1-2F)(2F+2)}{\Delta_\perp + \zeta_{-1}} \begin{Bmatrix} J & J' & 1 \\ F-1 & F & I \end{Bmatrix}^2 = 0\label{appeq:d2perpcond}.
\end{equation}
This can be rearranged into a quadratic equation in the detuning. It has two solutions of the form
\begin{equation}
\Delta_\perp = \frac{-b \pm \sqrt{b^2 - 4ac}}{2a},
\end{equation}
where
\begin{subequations}
\begin{align}
a &= 2F(2F+3) \begin{Bmatrix} J & J' & 1 \\ F+1 & F & I \end{Bmatrix}^2 - (4F+2)\begin{Bmatrix} J & J' & 1 \\ F & F & I \end{Bmatrix}^2 + (1-2F)(2F+2) \begin{Bmatrix} J & J' & 1 \\ F-1 & F & I \end{Bmatrix}^2, \label{appeq:quada}\\
b &= \zeta_{-1}2F(2F+3) \begin{Bmatrix} J & J' & 1 \\ F+1 & F & I \end{Bmatrix}^2 - (\zeta_{-1}+\zeta_{+1})(4F+2)\begin{Bmatrix} J & J' & 1 \\ F & F & I \end{Bmatrix}^2 + \zeta_{+1}(1-2F)(2F+2) \begin{Bmatrix} J & J' & 1 \\ F-1 & F & I \end{Bmatrix}^2, \\
c &= - (4F+2)\begin{Bmatrix} J & J' & 1 \\ F & F & I \end{Bmatrix}^2\zeta_{-1}\zeta_{+1}.\label{appeq:quadc}
\end{align}\label{appeq:quad}
\end{subequations}
\noindent This is guaranteed to yield two solutions because $a\neq 0$ and $b^2 - 4ac > 0$ for $F\geq1$, which is required to scatter off the $D_2$ line, as shown below.

\subsection{Proof that there are always two roots to the equation for $\Delta_\perp$ for scattering on the $D_2$ line~\label{app:realroots}}

Here we demonstrate the presence of two values of $\Delta_\perp$ that solve Eq.~\eqref{appeq:d2perpcond} for the cases of $J=1/2$ and $J'=3/2$. We define
\begin{subequations}
\begin{align}
A &= 2F(2F+3) \begin{Bmatrix} \frac{1}{2} & \frac{3}{2} & 1 \\ F+1 & F & I \end{Bmatrix}^2, \\
B &= - (4F+2)\begin{Bmatrix} \frac{1}{2} & \frac{3}{2} & 1 \\ F & F & I \end{Bmatrix}^2, \\
C &= (1-2F)(2F+2) \begin{Bmatrix} \frac{1}{2} & \frac{3}{2} & 1 \\ F-1 & F & I \end{Bmatrix}^2,
\end{align}
\end{subequations}
such that the coefficients of the quadratic equation read as
\begin{equation}
a = A + B + C, \;\;\;\;\;\; b = \zeta_{-1}A + (\zeta_{-1} + \zeta_{+1})B + \zeta_{+1}C, \;\;\;\;\;\; c = \zeta_{-1}\zeta_{+1}B.
\end{equation}
For there to be two physical roots we require $a \neq 0$ and $b^2-4ac > 0$. For the cases where $J=1/2$ and $J'=3/2$, we can use the identity
\begin{equation}
\frac{F^2(F+J-I+2)(F+J+I+3)}{(J+I-F+1)(F+I-J)(2F+3)} - 2F - 1 + \frac{(F+1)^2(J+I-F+2)(F+I-J-1)}{(F+J-I+1)(2F-1)(F+J+I+2)} = 0
\end{equation}
which will be proven in Appendix~\ref{app:importantzero}. We then simplify the expression such that the condition for $a=0$ is
\begin{equation}
\begin{Bmatrix} \frac{1}{2} & \frac{3}{2} & 1 \\ F+1 & F & I \end{Bmatrix}^2 = \begin{Bmatrix} \frac{1}{2} & \frac{3}{2} & 1 \\ F & F & I \end{Bmatrix}^2.
\end{equation}
This cannot be true for any combination of $F$ and $I$ that satisfies $F=I\pm J$ and $F\geq 1$, and so there are always two roots to the quadratic equation. For those roots to be real we require
\begin{equation}
(A\zeta_{-1} + B(\zeta_{-1} + \zeta_{+1}) + C\zeta_{+1})^2 > 4(A+B+C)(B\zeta_{-1}\zeta_{+1}).
\end{equation}
Expanding out allows for the rearrangement
\begin{equation}
 A^2\zeta_{-1}^2 + 2AC\zeta_{-1}\zeta_{+1} + B^2(\zeta_{-1} - \zeta_{+1})^2  + C^2\zeta_{+1}^2 > 2AB\zeta_{-1}(\zeta_{+1} - \zeta_{-1}) + 2BC\zeta_{+1}(\zeta_{-1} - \zeta_{+1}).
\end{equation}
We can use the expressions above to see that $A$ is positive, $B$ is negative, and $C$ is negative for $F\geq 1$ which is required for scattering on the $D_2$ line. Furthermore, $\zeta_{\pm1}$ have opposite signs. This means that all terms on the left-hand side are positive, while only one of the terms on the right-hand side can be, depending on the sign of $\zeta_{+1} - \zeta_{-1}$. We can bound the size of these terms through (the same process can be followed if $2BC\zeta_{+1}(\zeta_{-1} - \zeta_{+1})$ is positive)
\begin{gather}
\left(A\zeta_{-1} - B\left(\zeta_{+1}-\zeta_{-1}\right)\right)^2 \geq 0 
\Rightarrow A^2\zeta_{-1} + B^2\left(\zeta_{+1}-\zeta_{-1}\right)^2 \geq 2AB\zeta_{-1}(\zeta_{+1} - \zeta_{-1}),
\end{gather}
and show that the inequality is always valid, and Eq.~\eqref{appeq:d2perpcond} always has two real non-trivial roots.

\section{Derivation of $\Delta_\parallel$\label{app:dpar}}

In Sec.~\ref{sec:amplitude}, we find a detuning where
\begin{equation}
   S_{m,F,F'}(\omega) = \sum\limits_{F'} \frac{r^2_{m,F,F'} + s^2_{m,F,F'}}{\Delta_{F'}}
\end{equation}
is independent of $m$. By substituting in expressions for the dipole matrix elements and Wigner 3-j symbols (see Appendix~\ref{app:dme}) we find
\begin{align}
S_{m,F,F'}(\omega) = \sum\limits_{F'}\frac{1}{\Delta_{F'}}\frac{2(2F'+1) (F+F'-1)!\begin{Bmatrix} J & J' & 1 \\ F' & F & I \end{Bmatrix}^2}{(F'-F+1)!(F-F'+1)!(F+F'+2)!} \left[\frac{(F - m)! (F' + m+1)!}{(F + m)!(F' - m-1)!} + \frac{(F + m)! (F'- m+1)!}{(F- m)!(F'+ m-1)!}\right].\label{eq:linearamplitudecondition}
\end{align} 
This can be simplified to
\begin{equation}
S_{m,F,F'}(\omega) = \left[\frac{F^2 + 3F + 2 + m^2}{(2F+1)(2F+2)}\frac{\begin{Bmatrix} J & J' & 1 \\ F+1 & F & I \end{Bmatrix}^2}{\Delta + \zeta_{+1}} 
+ \frac{F^2+F-m^2}{F(2F+2)} \frac{\begin{Bmatrix} J & J' & 1 \\ F & F & I \end{Bmatrix}^2}{\Delta}
+ \frac{F^2-F+m^2}{2F(2F+1)}\frac{ \begin{Bmatrix} J & J' & 1 \\ F-1 & F & I \end{Bmatrix}^2}{\Delta+\zeta_{-1}}\right],
\end{equation} 
where we use the same convention above that detunings are defined from the $F'=F$ level. For the scattering to be $m$ independent simply requires that the sum of the quadratic terms is zero. The detuning $\Delta_\parallel$ thus satisfies the condition
\begin{equation}
\frac{2F}{\Delta_\parallel + \zeta_{+1}} \begin{Bmatrix} J & J' & 1 \\ F+1 & F & I \end{Bmatrix}^2
- \frac{4F+2}{\Delta_\parallel} \begin{Bmatrix} J & J' & 1 \\ F & F & I \end{Bmatrix}^2
+ \frac{2F+2}{\Delta_\parallel+\zeta_{-1}} \begin{Bmatrix} J & J' & 1 \\ F-1 & F & I \end{Bmatrix}^2 = 0.\label{appEq:d2ampcondition}
\end{equation}
As above, this can be rearranged to a quadratic equation in $\Delta_\parallel$. However, for scattering on the $D_2$ line, the $\Delta_\parallel^2$ coefficient in that equation is 
\begin{equation}
2F \begin{Bmatrix} \frac{1}{2} & \frac{3}{2} & 1 \\ F+1 & F & I \end{Bmatrix}^2 - (4F+2) \begin{Bmatrix} \frac{1}{2} & \frac{3}{2} & 1 \\ F & F & I \end{Bmatrix}^2 + (2F+2) \begin{Bmatrix} \frac{1}{2} & \frac{3}{2} & 1 \\ F-1 & F & I \end{Bmatrix}^2 = 0\label{eq:importantzero}.
\end{equation}
This is also true for other combinations of $J$, $J'$ and $F$, but not all, as discussed in Appendix~\ref{app:importantzero}. This means that Eq.~\eqref{appEq:d2ampcondition} has only one solution
\begin{equation}
 \Delta_\parallel = \frac{(2F+1)\begin{Bmatrix} \frac{1}{2} & \frac{3}{2} & 1 \\ F & F & I \end{Bmatrix}^2 \zeta_{+1}\zeta_{-1}}{F\begin{Bmatrix} \frac{1}{2} & \frac{3}{2} & 1 \\ F+1 & F & I \end{Bmatrix}^2 \zeta_{-1} - (2F+1)\begin{Bmatrix} \frac{1}{2} & \frac{3}{2} & 1 \\ F & F & I \end{Bmatrix}^2(\zeta_{+1} + \zeta_{-1}) + (F+1)\begin{Bmatrix} \frac{1}{2} & \frac{3}{2} & 1 \\ F-1 & F & I \end{Bmatrix}^2\zeta_{+1}}.\label{eq:ampsolutiond2}
 \end{equation}
 
\subsection{Proof that there is only one root to the equation for $\Delta_\parallel$ for {scattering on the $D_2$ line}\label{app:importantzero}}

The quadratic coefficient for the condition to find $\Delta_\parallel$ is
\begin{equation}
D = 2F \begin{Bmatrix} J & J' & 1 \\ F+1 & F & I \end{Bmatrix}^2 - (4F+2) \begin{Bmatrix} J & J' & 1 \\ F & F & I \end{Bmatrix}^2 + (2F+2) \begin{Bmatrix} J & J' & 1 \\ F-1 & F & I \end{Bmatrix}^2.~\label{eq:quadraticzero}
\end{equation}
For there to be a single non-trivial root for $\Delta_\parallel$, and for the magic detuning to exist when the hyperfine splitting is dominated by the magnetic dipole contribution, we require $D=0$. There are three possible cases of $J'=\{J-1,J,J+1\}$. In the following, we treat the specific case of $J'=J+1$, a class which includes scattering on a $D_2$ line. In this case we have that
\begin{equation}
\begin{Bmatrix} J & J+1 & 1 \\ F' & F & I \end{Bmatrix}^2 = g(F,I,J) \frac{(F'+J-I+1)!(J+I-F'+1)!(F'+F-1)!(F'+J+I+2)!}{(F'+I-J-1)!(F'-F+1)!(F-F'+1)!(F'+F+2)!},
\end{equation}
where $g(F,I,J)$ is a function that does not depend on $F'$ and so is a constant factor in Eq.~\eqref{eq:quadraticzero}. Using this expression we find that the condition for $D=0$ becomes
\begin{equation}
\frac{F^2(F+J-I+2)(F+J+I+3)}{(J+I-F+1)(F+I-J)(2F+3)} - 2F - 1 + \frac{(F+1)^2(J+I-F+2)(F+I-J-1)}{(F+J-I+1)(2F-1)(F+J+I+2)} = 0.\label{eq:keycondition}
\end{equation}

The ground state hyperfine level must have $F = I + J - n$ where $n \in \mathbb{N}$ is a positive integer (or zero) such that $F\geq1$, $F\geq J-I$ and $F\geq I-J$. We can use this to simplify the expression and find solutions for a particular $n$. A few pertinent examples are given below. Cases not covered here can be found by setting $F=I+J-n$, inputting in the specific values of $I$ and $J$ and solving for $n$. If $n$ meets the criteria above then a magic detuning exists for that combination of $F$, $I$ and $J$.

\subsubsection{$n=0$}

For $n=0$, $F=I+J$ is the maximum angular momentum hyperfine level. Solutions are found by solving
\begin{equation}
F^2(2F-1)(2J+2)(2J+1) - (2F+1)(2F-1)(2F-2J)(2J+1) + (F+1)(2F-2J-1)(2F-2J) = 0.
\end{equation}
This expression has five roots which can be found by rearranging to the form
\begin{equation}
2F(2F^2+3F+1)J(2J-1) = 0.
\end{equation}
Three of the roots are found as $F=0,-1/2,-1$. All of these are unphysical as we require $F\geq 1$. As such, the only roots are $J=0,1/2$. This means a magic detuning exists for scattering via a line where $J'=J+1$ for $J=0$, $F=I$, and for $J=1/2$, $F=I+1/2$. The second case is exactly that for scattering on the $D_2$ line from the highest $F$ hyperfine level.

\subsubsection{$n=1$}

For $n=1$, $F=I+J-1$ is the second highest angular momentum hyperfine level. Solutions are found by solving
\begin{equation}
F^2J(2F-1)(2J+1)(F+2) - J(2F+1)(2F-1)(2F-2J+1)(2F+3) + 3(F+1)^2(F-J)(2F-2J+1) = 0.
\end{equation}
Unlike the equation above, this does not reduce to an easily solved form, but it can be shown that there are physical solutions for $J=1/2$ with $F$ free and for $F=5, J=2$. The former case is again that for the $D_2$ line, but now from the lower hyperfine level.

\section{Derivation of $\Delta_\perp = \Delta_\parallel$ when the hyperfine splitting is dominated by the magnetic dipole contribution\label{app:hype}}

Here, we consider a hyperfine splitting given solely by the magnetic dipole moment, i.e.,
\begin{equation}
\delta_{\mathrm{hfs}} = \ahfsf \left[F'(F'+1) - I(I+1) - J(J+1)\right]
\end{equation}
where $A_{\mathrm{hfs}}$ is the magnetic dipole hyperfine constant. This yields
\begin{subequations}
 \begin{gather}
 \zeta_{+1} = -A_{\mathrm{hfs}}(F+1), \\
 \zeta_{-1} = A_{\mathrm{hfs}}F.
 \end{gather}
\end{subequations}
By substituting the approximations of the hyperfine splitting above into the detuning solutions, we show that all three solutions coincide. An important part of this derivation harnesses that Eq.~\ref{eq:importantzero} is satisfied. While we consider $J=1/2$ and $J'=3/2$ here, it should be noted that the following exists for all $J,J'$ where Eq.~\ref{eq:importantzero} is true.

\subsection{$\Delta_\perp$ when the hyperfine splitting is dominated by the magnetic dipole contribution}

Using the combination of Wigner 6-j coefficients in Eq.~\eqref{eq:importantzero} to eliminate the $F'=F-1$ terms, we arrive at the following form for the quadratic coefficients in Eq.~\eqref{appeq:quad}:
\begin{subequations}
\begin{align}
\frac{a}{\ahfs} &= 4F(2F+1) \left[\begin{Bmatrix} \frac{1}{2} & \frac{3}{2} & 1 \\ F+1 & F & I \end{Bmatrix}^2 - \begin{Bmatrix} \frac{1}{2} & \frac{3}{2} & 1 \\ F & F & I \end{Bmatrix}^2 \right], \\
\frac{b}{\ahfs} &= 2F(2F+1) \begin{Bmatrix} \frac{1}{2} & \frac{3}{2} & 1 \\ F+1 & F & I \end{Bmatrix}^2 + 2F(2F+1)^2\begin{Bmatrix} \frac{1}{2} & \frac{3}{2} & 1 \\ F & F & I \end{Bmatrix}^2, \\
\frac{c}{\ahfs} &= F(4F+2)(F+1)\begin{Bmatrix} \frac{1}{2} & \frac{3}{2} & 1 \\ F & F & I \end{Bmatrix}^2.
\end{align}
\end{subequations}
One can then show that
\begin{equation}
\frac{b^2-4ac}{\ahfs^2} = \left[2F(2F+1)\begin{Bmatrix} \frac{1}{2} & \frac{3}{2} & 1 \\ F+1 & F & I \end{Bmatrix}^2 - 2F(2F+1)(2F+3)\begin{Bmatrix} \frac{1}{2} & \frac{3}{2} & 1 \\ F & F & I \end{Bmatrix}^2\right]^2
\end{equation}
and make the simplification
\begin{equation}
\frac{\Delta_\perp}{\ahfs} = \frac{-\begin{Bmatrix} \frac{1}{2} & \frac{3}{2} & 1 \\ F+1 & F & I \end{Bmatrix}^2 - (2F+1)\begin{Bmatrix} \frac{1}{2} & \frac{3}{2} & 1 \\ F & F & I \end{Bmatrix}^2 \pm \left[\begin{Bmatrix} \frac{1}{2} & \frac{3}{2} & 1 \\ F+1 & F & I \end{Bmatrix}^2 - (2F+3)\begin{Bmatrix} \frac{1}{2} & \frac{3}{2} & 1 \\ F & F & I \end{Bmatrix}^2\right]}{4\left[\begin{Bmatrix} \frac{1}{2} & \frac{3}{2} & 1 \\ F+1 & F & I \end{Bmatrix}^2 - \begin{Bmatrix} \frac{1}{2} & \frac{3}{2} & 1 \\ F & F & I \end{Bmatrix}^2 \right]},
\end{equation}
such that the two solutions are
\begin{subequations}
\begin{align}
\frac{\Delta_\perp}{\ahfs} &= \frac{-(F+1)\begin{Bmatrix} \frac{1}{2} & \frac{3}{2} & 1 \\ F & F & I \end{Bmatrix}^2}{\begin{Bmatrix} \frac{1}{2} & \frac{3}{2} & 1 \\ F+1 & F & I \end{Bmatrix}^2 - \begin{Bmatrix} \frac{1}{2} & \frac{3}{2} & 1 \\ F & F & I \end{Bmatrix}^2}, \\
\frac{\Delta_\perp}{\ahfs} &= \frac{1}{2}.
\end{align}    
\end{subequations}
One of these coincides exactly with $\Delta_\pi$.

\subsection{$\Delta_\parallel$ when the hyperfine splitting is dominated by the magnetic dipole contribution}

The same trick can be used to simplify the denominator of Eq.~\eqref{eq:ampsolutiond2}. Leveraging the zero combination of the Wigner 6-j coefficients and substituting in the approximations to the hyperfine splitting, one arrives at
\begin{equation}
\frac{\Delta_\parallel}{\ahfs} = \frac{-(F+1)\begin{Bmatrix} \frac{1}{2} & \frac{3}{2} & 1 \\ F & F & I \end{Bmatrix}^2}{\begin{Bmatrix} \frac{1}{2} & \frac{3}{2} & 1 \\ F+1 & F & I \end{Bmatrix}^2 - \begin{Bmatrix} \frac{1}{2} & \frac{3}{2} & 1 \\ F & F & I \end{Bmatrix}^2}.
\end{equation}
This matches $\Delta_\pi$, and thus one of the solutions $\Delta_\perp$, and thus confirms the existence of a magic detuning if the hyperfine splitting can be approximated as arising solely from magnetic dipole moment.

\section{Considering the $D_1$ line\label{app:d1line}}

We can follow the same procedure to find detunings that meet our conditions for scattering on the $D_1$ line, where scattering is via two levels with $F' \in \{F,F+1\}$ if $F=I-1/2$ and $F' \in \{F-1,F\}$ if $F=I+1/2$. However, the values of $\Delta_\perp$ and $\Delta_\parallel$ are never equal, and so a magic detuning cannot be found.

\subsection{Finding $\Delta_\perp$}

We now find a single solution to Eq.~\eqref{eq:d2ellipticitycondition} as one of the terms has a zero Wigner 6j coefficient. As above, we define $\Delta_\perp$ as the detuning at which ellipticity goes to zero. We find
\begin{subequations}
 \begin{align}
 F=I-\frac{1}{2}: ~~~\Delta_\perp &= \frac{\zeta_{+1}(2F+1)\begin{Bmatrix} J & J' & 1 \\ F & F & I \end{Bmatrix}^2}{F(2F+3)\begin{Bmatrix} J & J' & 1 \\ F+1 & F & I \end{Bmatrix}^2 - (2F+1) \begin{Bmatrix} J & J' & 1 \\ F & F & I \end{Bmatrix}^2}, \\
 F=I+\frac{1}{2}:~~~\Delta_\perp &= \frac{\zeta_{-1}(2F+1)\begin{Bmatrix} J & J' & 1 \\ F & F & I \end{Bmatrix}^2}{(1-2F)(F+1)\begin{Bmatrix} J & J' & 1 \\ F-1 & F & I \end{Bmatrix}^2 - (2F+1) \begin{Bmatrix} J & J' & 1 \\ F & F & I \end{Bmatrix}^2}.
 \end{align}
\end{subequations}

\subsection{Finding $\Delta_\parallel$}

For scattering on the $D_1$ line via two levels with $F' \in \{F,F\pm1\}$, the detunings are
\begin{subequations}
 \begin{align}
 F=I-\frac{1}{2}: ~~~ \Delta_\parallel &= \frac{\zeta_{+1}(4F+2)\begin{Bmatrix} J & J' & 1 \\ F & F & I \end{Bmatrix}^2}{2F\begin{Bmatrix} J & J' & 1 \\ F+1 & F & I \end{Bmatrix}^2 - (2F+1)\begin{Bmatrix} J & J' & 1 \\ F & F & I \end{Bmatrix}^2}, \\
 F=I+\frac{1}{2}: ~~~ \Delta_\parallel &= \frac{\zeta_{-1}(4F+2)\begin{Bmatrix} J & J' & 1 \\ F & F & I \end{Bmatrix}^2}{(2F+2)\begin{Bmatrix} J & J' & 1 \\ F-1 & F & I \end{Bmatrix}^2 - (2F+1)\begin{Bmatrix} J & J' & 1 \\ F & F & I \end{Bmatrix}^2}
 \end{align}
\end{subequations}

\section{Effective Hamiltonian treatment for cavity QED \label{app:cavity}}

\begin{figure*}[t!]
\includegraphics[width=0.75\textwidth]{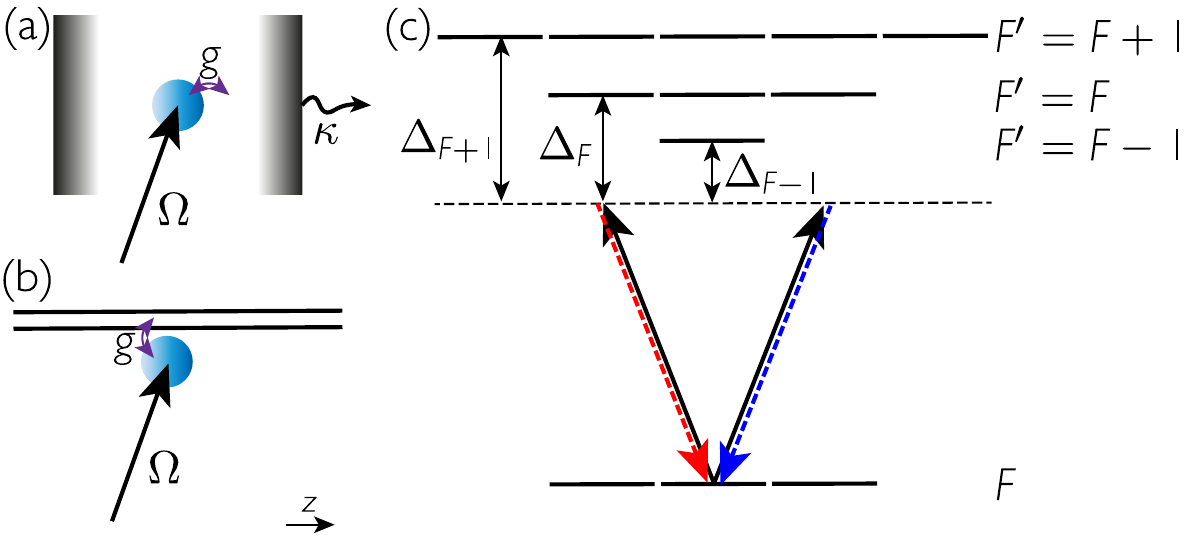}
\caption{Setup. A single atom coupled strongly to (a) a cavity or (b) a waveguide is driven off resonantly at a Rabi frequency $\Omega$. The atom is coupled to the cavity or waveguide at a strength $g$, such that light is scattered by the atom off three detuned hyperfine levels into the cavity and waveguide. The cavity axis or waveguide is chosen as the quantization axis $z$, such that the drive and cavity and waveguide modes couple to $\sigma_\pm$-polarized transitions in the atom.\label{fig:cavitysetup}}
\end{figure*}

As in the main text, we consider an atom initialized in a hyperfine state $F$, with $2F+1$ {Zeeman states} $\set{-m,-m+1,\dots,m}$ and driven on the $D_2$ line by an arbitrary superposition of $\sigma_\pm$-polarized light. The atom can scatter into a cavity (or equivalently a waveguide) [see Figure~\ref{fig:cavitysetup}]. The cavity has two modes with $\sigma_\pm$ polarization and degenerate frequency.

The system follows a master equation ($\hbar=1$)
\begin{equation}
\dot{\rho} = -\ii [\hat{H},\rho] + \kappa\mathcal{L}[\hat{a}]\rho + \kappa\mathcal{L}[\hat{b}]\rho
\end{equation}
where  $\hat{a}~(\hat{b})$ is the annihilation operator for the $\sigma_+$~($\sigma_-$) polarized cavity mode, $\mathcal{L}[\hat{O}]\rho = 2\hat{O}\rho\hat{O}^\dagger - \hat{O}^\dagger\hat{O}\rho - \rho\hat{O}^\dagger\hat{O}$ and the total Hamiltonian is $\hat{H} = \hat{H}_{\mathrm{sys.}} + \hat{H}_{\mathrm{\mathrm{int.}}}$. In the rotating frame of the drive, the system Hamiltonian is
\begin{align}
\hat{H}_{\mathrm{sys.}} &= \delta \left(\hat{a}^\dagger\hat{a} + \hat{b}^\dagger\hat{b}\right) + \sum\limits_{F'} \Delta_{F'} \sum\limits_{m_{F'}} \ket{F',m_{F'}}\bra{F',m_{F'}},
\end{align}
where $\delta = \omega - \omega_c$ is the detuning between drive and cavity mode, $\Delta_{F'} = \omega - \omega_{F'}$ is the detuning of the drive from the excited manifold $F'$. We assume that all Zeeman shifts are negligible such that all {Zeeman states} can be considered at the same energy. The interaction Hamiltonian is
\begin{equation}
\hat{H}_{\mathrm{int.}} = \sum\limits_{F,F'} \left(\frac{\cos\theta~\Omega}{2} + g \hat{a}\right) \sum\limits_{m} r_{m,F,F'}\ket{F',m+1}\bra{F,m} + \sum\limits_{F,F'} \left(\frac{\sin\theta~\Omega}{2} + g \hat{b}\right) \sum\limits_{m} s_{m,F,F'}\ket{F',m-1}\bra{F,m} + \mathrm{H.c.},
\end{equation}
where $\Omega$ is the Rabi frequency of the drive, $g$ is the coupling strength of the atom to the cavity and $\theta$ encodes the ellipticity of the drive, as in the main text.

In the limit of $\Omega / \Delta_{F'} \ll 1~\forall~F'$, the atomic saturation is essentially zero, and one can adiabatically eliminate the excited states. This yields an effective Hamiltonian
\begin{align}
\hat{H}_\mathrm{eff.} &= \delta (\hat{a}^\dagger \hat{a} + \hat{b}^\dagger\hat{b}) + \sum\limits_{m} \left[\left(U_{m}^a \hat{a}^\dagger\hat{a} + U_{m}^b \hat{b}^\dagger\hat{b} + \omega_{m}\right)\ket{m}\bra{m} + \tilde{\Omega}_{m} \left(\ket{m+2}\bra{m} + \ket{m}\bra{m+2}\right) \right.\notag\\& + h_{m} \left(\hat{b}^\dagger \hat{a} \ket{m+2}\bra{m} + \hat{a}^\dagger\hat{b}\ket{m}\bra{m+2}\right)+ \eta^{+,\mathrm{Ray.}}_{m} \left(\hat{a} + \hat{a}^\dagger\right)\ket{m}\bra{m}
+  \eta^{-,\mathrm{Ray.}}_{m} \left(\hat{b} + \hat{b}^\dagger\right)\ket{m}\bra{m} \notag\\&+ \left. \cos\theta~\eta^{\mathrm{Ram.}}_{m} \left( \hat{b}^\dagger \ket{m+2}\bra{m} + \hat{b} \ket{m}\bra{m+2} \right) +  \sin\theta~\eta^{\mathrm{Ram.}}_{m} \left( \hat{a}^\dagger \ket{m}\bra{m+2} + \hat{a} \ket{m+2}\bra{m} \right)\right],\label{eq:hamiltonianeff}
\end{align}
where we have dropped the $F$ notation from the atomic states and defined various effective parameters. The effective frequency of the atomic states is given by the light shift from the drive laser and from cavity population
\begin{gather}
U_{m}^{a} = \sum\limits_{F'} \frac{g^2 \rtwo}{\Delta_{F'}}, \;\;\;\; U_{m}^{b} = \sum\limits_{F'} \frac{g^2 \stwo}{\Delta_{F'}}, \;\;\;\;
\omega_{m} = \sum\limits_{F'} \frac{\Omega^2\left(\cos^2\theta~\rtwo + \sin^2\theta~\stwo\right)}{4\Delta_{F'}}.
\end{gather}
There are also coupling terms between atomic states which conserve photon number. This corresponds to two-photon transitions mediated by two oppositely-polarized drive photons or the transfer of population between the two cavity modes. These processes occur at rates
    \begin{gather}
        \tilde{\Omega}_{m} = \sum\limits_{F'} \frac{\cos\theta~\sin\theta~\Omega^2\rscross}{4\Delta_{F'}}, \;\;\;\;
        h_{m} = \sum\limits_{F'} \frac{g^2\rscross}{\Delta_{F'}}.
    \end{gather}
The final group of terms correspond to describe processes where the atoms scatter into the cavity. This can happen in two ways: returning to the same atomic state or changing atomic state. Employing the same designations of Rayleigh and Raman scattering as in the main text, these scattering rates read
    \begin{gather}
\eta^{+,\mathrm{Ray.}}_{m} = \sum\limits_{F'} \frac{\cos\theta~\Omega g\rtwo}{2\Delta_{F'}}, \;\;\;\;
\eta^{-,\mathrm{Ray.}}_{m} = \sum\limits_{F'} \frac{\sin\theta~\Omega g\stwo}{2\Delta_{F'}}, \;\;\;\;
\eta^{\mathrm{Ram.}}_{m} = \sum\limits_{F'} \frac{\Omega g\rscross}{2\Delta_{F'}}.
    \end{gather}

\subsection{Simplifications at the magic detuning}

It is convenient to transform the effective Hamiltonian into the parallel and perpendicular basis. These modes are composed of the circularly polarized modes as
\begin{gather}
    \hat{c}_\parallel  = \cos\theta~\hat{a} + \sin\theta~\hat{b}, \;\;\;\;\;\;
\hat{c}_\perp = \sin\theta~\hat{a} - \cos\theta~\hat{b}.
\end{gather}
Using these expressions and assuming we operate at a magic detuning, the above Hamiltonian reads as
\begin{align}
    H_\mathrm{eff.} &= \delta \left(\cpar^\dagger\cpar + \cperp^\dagger\cperp\right) + \sum\limits_{m} \left[ \left(U^{\parallel} \cpar^\dagger\cpar + U^{\perp} \cperp^\dagger \cperp + \omega_{m} \right) \ket{m}\bra{m} +\eta^{\parallel,\mathrm{Ray.}} \left(\cpar + \cpar^\dagger\right)\ket{m}\bra{m} \right] \\
    &= \left(\delta + U^{\parallel}\right)\cpar^\dagger\cpar + \left(\delta + U^{\perp}\right)\cperp^\dagger\cperp + \eta^{\parallel,\mathrm{Ray.}} \left(\cpar + \cpar^\dagger\right),
\end{align}
where we have defined new $m$-independent dispersive shifts and cavity scattering rates. These can be calculated for any choice of $m$ and take the form
\begin{subequations}
    \begin{gather}
        U^{\parallel} = \cos^2\theta~U^{a}_{m} + \sin\theta^2~U^b_{m}, \;\;\;\;
        U^{\perp} = \sin^2\theta~U^{a}_{m} + \cos^2\theta~U^b_{m}, \;\;\;\; \eta^{\parallel,\mathrm{Ray.}} = \cos\theta~\eta^{+,\mathrm{Ray.}}_{m} + \sin\theta~\eta^{-,\mathrm{Ray.}}_{m}.
    \end{gather}
\end{subequations}
In this form, the Hamiltonian reduces to that of an off-resonantly driven cavity. While the perpendicular cavity mode is also dispersively shifted, it does not play a role in dynamics. We thus write the single atom Hamiltonian as
\begin{equation}
  H_{\mathrm{eff.}} = \left(\delta + U^{\parallel}\right)\cpar^\dagger\cpar + \eta^{\parallel,\mathrm{Ray.}} \left(\cpar + \cpar^\dagger\right).
\end{equation}

We now consider $N$ atoms, each coupled to the cavity at $g_j = \sin\varphi_j g_0$, where $\varphi_j = \mathbf{k}_c\cdot\mathbf{r}_j$ encodes the strength of coupling between an atom at position $\mathbf{r}_j$ to a standing wave cavity mode with wavenumber $\mathbf{k}_c$. The Hamiltonian is then the sum over these $N$ atoms and gives
\begin{equation}
   H_{\mathrm{eff.}} = \delta \cpar^\dagger\cpar + \sum\limits_{j=1}^N \sin^2\varphi_j~U^{\parallel}\cpar^\dagger\cpar + \sin\varphi_j~\eta^{\parallel,\mathrm{Ray.}} \left(\cpar + \cpar^\dagger\right). 
\end{equation}
If all atoms couple equally and in phase then $\sum_j\sin\varphi_j = N$, and the field intensity scales as $N^2$. If atoms couple equally but with alternating signs, then the sum is zero for an even number of atoms or one for an odd number. These results align with those observed in Ref.~\cite{Zhenjie23}.

\end{document}